# Effects of stochastic and natural seismic noise on the performance of waveform cross-correlation used to recover low-magnitude seismicity prior to the July 29, 2025, Kamchatka earthquake

Ivan O. Kitov

**Abstract**

Waveform cross-correlation (WCC) applied to data from seismic arrays allows for the reduction of the detection threshold by approximately an order of magnitude. When applied to the data of the International Monitoring System, the smallest WCC detected events were by at least one magnitude unit lower than those reported by the International Data Centre (IDC) and revealed that the pattern of low-magnitude activity prior to the July 29, 2025, earthquake was similar to that prior to the May 24, 2013, Sea of Okhotsk earthquake. The consistent increase in the number of events with magnitudes approaching the corner value of the Kamchatka recurrence curve can likely be used as a precursory indicator of a mega-earthquake preparation. The WCC-based events are characterized by detections with a signal-to-noise ratio below the IDC detection threshold and are often not visible to analysts. This makes the statistical significance of the WCC event hypotheses uncertain due to the absence of a random noise reference and the potential side-sensitivity of the IMS arrays to high-amplitude signals from remote sources. Both problems are addressed.

A computer-generated random noise is used to calculate a WCC bulletin according to the same procedure as was used for the actual data. This exercise shows a negligibly low number of WCC events generated by random-noise waveforms and confirms the statistical significance of the WCC-based events. The side-sensitivity is modelled using the March 11, 2011, Tohoku earthquake. When only master events (MEs) within the Kamchatka region are used in the WCC processing, a large number of false events are generated. When MEs from the Tohoku zone are added, all false events disappear since the Tohoku MEs win the conflict resolution process. The random noise added to the actual data can suppress false events created by the Kamchatka MEs from Tohoku signals, but it also suppresses valid events in the Kamchatka region. When the random noise is scaled to a small fraction of the maximum amplitude within the processed interval, the WCC detection is enhanced with more detections and WCC-based events created. This is an effect similar to noise "whitening", which makes the matched filter detector more sensitive and closer to its optimal performance under stochastic noise conditions. In the subsequent study, this stochastic noise effect will be used to improve the WCC bulletins for the Kamchatka and Sea of Okhotsk mega-earthquakes. A set of optimal random noise scaling factors was estimated for further investigations.

## Introduction

Waveform cross-correlation (WCC) is an effective method to reduce the detection threshold in various seismological tasks known since the beginning of the progress of digital seismology [Israelsson, 1990; Joswig, 1990]. The gain in the number of detected events can reach an order of magnitude for regional and teleseismic networks [Schaff and Richards, 2004]. Another important feature provided by WCC is the possibility of extremely accurate measurements of the relative arrival time as achieved by synchronization of a template and sought signals. The accurate arrival times are converted into a high location accuracy of the sought events relative to the master event (ME), which is the source of signals detected at the stations associated with this ME [Waldhauser and Schaff, 2008; Selby, 2010]. These features lead to significant improvements in the completeness, consistency, and accuracy of seismic catalogs and bulletins [Bobrov *et al*,. 2014, 2016ab, 2017].

The WCC detector requires high-quality waveform templates. Both actual signals and synthetic seismograms [Bobrov *et al*., 2016b] can be used as templates, and the latter are more efficient in some cases [Kitov *et al*., 2016]. Between actual waveforms and synthetics lie semi-synthetic templates generated from a large number of actual records processed by various mathematical methods, from a simple SVD to modern AI. The areas with intense seismicity, such as the Kamchatka region, provide a sufficient number of repeating events to obtain the best possible set of representative MEs and their templates.

Seismic phased arrays are like the magnifying glass in the era of van Leeuwenhoek that allows the visualization of signals not detectable by three-component (3-C) stations. On average, the gain in detection threshold for an array is proportional to the square root of the number of individual sensors distributed within some area on the surface [Schweitzer *et al*., 2012]. Arrays are designed to magnify a signal, i.e. to increase the signal-to-noise ratio (SNR), by noise suppression via the destructive interference when the signals from individual channels are stacked. Time delays between arrivals at individual channels for a given plane wave with a known apparent velocity vector can be calculated from geographical positions of the sensors.

The accuracy of the time-delay prediction in the beamforming method depends on the difference between the theoretical and the actual plane wave's apparent velocity. The larger the difference, the higher the beam-loss, i.e., the power of the signal obtained with theoretical arrival times relative to that which could be obtained with the actual arrival times [Schweitzer *et al*., 2012]. The WCC method applied to seismic arrays possesses not only the advantage of the matched filter detector [Turin, 1960] but also suffers no beam loss effect, as the sought and template signals must be fully synchronized at individual sensors. Any deviation from absolute synchronization must be due to the difference between the master and sought events' relative positions, and, thus, can be accurately predicted across the channels.

The combination of WCC and arrays provides the most efficient detection method used for scientific research and applied purposes. The International Monitoring System (IMS) includes a global seismological network of 170 stations (50 primary and 120 auxiliary) in its final configuration under the Comprehensive Nuclear-Test-Ban Treaty (CTBT) provisions [CTBT, 1996]. Not all stations are in operation yet, but there are already 34 working arrays. They were successfully used in various studies from far-teleseismic to near-regional distances. One such study focused on the May 24, 2013, Sea of Okhotsk mega-earthquake demonstrated the capability of the WCC and arrays symbiosis in the detection of a large number of low-magnitude earthquakes immediately prior to the mainshock. These events were not detected by the International Data Centre (IDC) or any other seismological agency. At least, there are no records in the ISC database.

The evolution of low-magnitude seismicity before mega-earthquakes is a potential indicator of the approaching mainshock [Schaff *et al*., 2025]. Both seismic arrays and WCC are mandatory ingredients of a successful investigation into low-magnitude seismicity, but not all detected events can be confirmed visually by IDC analysts as they are mainly deep in the ambient noise. The WCC method can find very weak signals in line with the discrete Fourier transform estimating amplitudes of high frequency harmonics orders of magnitude lower than those of at lower frequencies. However, the event hypotheses built in the WCC pipeline [Kitov, 2026d] are then missing one of the principal sources of their statistical significance. In order to provide sufficient information on the probability of an event hypothesis created in the WCC pipeline of being valid, one has to present the case of random distribution of arrivals as a reference.

The ambient noise is not purely stochastic, as it created by a finite mixture of regular seismic phases from various sources. This makes any experiment with the ambient noise intended to represent a random process, limited in predictive power, as the sought signals for a given template can be hidden in this very noise. A computer generated random noise is a better choice, as it must fit several principal requirements [Press *et al*., 1986], but its actual properties are case-dependent. Stochastic noise generated by a computer program is a tool to replace or suppress any actual signal to an infinitesimal level. It is used in this study to mimic actual waveforms in the WCC pipeline to produce a reference bulletin to compare with the actual cross-correlation Standard Event List (XSEL), which is an automatic bulletin similar to Standard Event List 3 produced by the IDC using the same IMS data [Coyne *et al*., 2012].

However, one can use a random time series to suppress noise components coherent with the templates [Adushkin *et al*., 2025]. The gain can be significant for both a relatively high SNR and the weakest signals. This method was successfully applied to the WCC processing of the July 25, 2025, Kamchatka earthquake data during the hour following the mainshock [Kitov *et al*., 2026]. Several aftershocks were detected during the first ten minutes after the start of the coseismic phase. The first minutes after megathrust events are well-known as a still period without any aftershocks detected. The focused study of the WCC detector with added random noise has shown high efficiency in cases where the ambient noise and the template are coherent [Kitov, 2026c].

This study extends the previous work on low-magnitude seismicity before the July 29, 2025, Kamchatka earthquake [Kitov, 2026d] in three directions, all related to stochastic noise. The first part is devoted to the simulation of actual waveforms with random noise generated by the *ran3* program [Press *et al.,* 1986]. The results of this simulation are compared with the results obtained by the WCC pipeline for actual data. This comparison confirms the statistical significance of the XSEL events used to reveal the growing low-magnitude seismic activity potentially serving as a precursory variable for the earthquake prediction. The second part is devoted to the investigation of the side-sensitivity of the WCC applied to array stations in view of the possibility of some XSEL events be false, as they are created from phases generated by strong sources outside of the studied area. The random seismic noise is used here to vary the irrelevant signals' amplitude to estimate side-sensitivity. The third part encompasses the advantages of the positive stochastic noise effect on the WCC pipeline. This is a feasibility study to compare selected results of the previous work with new results obtained with the WCC enhanced by random noise. A thorough testing of various stochastic noise amplitudes had been accomplished before the optimal value was estimated.

## WCC pipeline with stochastic noise as an actual waveform

### *Master events*

The effect of stochastic noise on the WCC processing starts from the selection of master events (MEs) and signal templates at associated stations. There are thousands of potential MEs in the studied region. For example, Figure 1 shows 6628 REB events between July 20 and December 4, 2025. The 100 best MEs were selected by [Kitov, 2026d] from the entire REB set from 2001 to 2023. The selection process was based on the pair-wise cross-correlation of each event with all its neighbours. The events that detected the largest number of neighbours were selected for the MEs set. The selected MEs were distributed according to their positions to the nodes of a regular Global Grid with an average spacing of approximately 1.25° [Kitov *et al.*, 2016]. From this original global set consisting of approximately 42,000 MEs, the 100 best MEs were selected for the WCC processing of the July 29, 2025, Kamchatka earthquake.

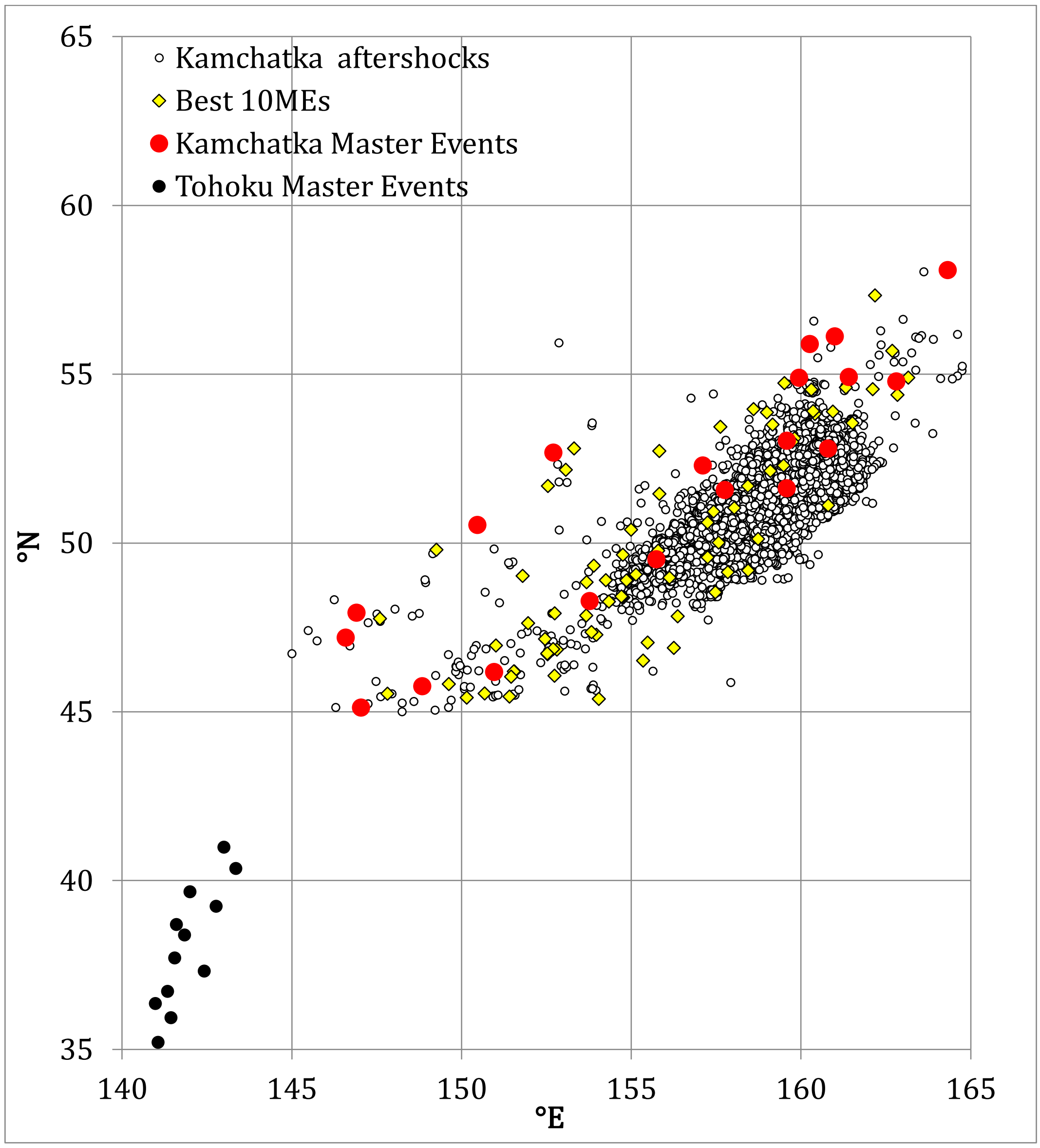

Figure 1. Selected master events for the Kamchatka (twenty MEs) and Tohoku (twelve MEs) regions. All 6,628 REB events in the Kamchatka aftershock sequence from July to December 2025 are plotted. The best 100 MEs are also shown to illustrate the distribution of the twenty MEs used in this study.

The results of the WCC processing revealed an increase in the low-magnitude earthquakes activity prior to the mainshock, which is interpreted as a precursor to the coseismic phase. One of the questions raised in [Kitov, 2026d] was the statistical significance of the event hypotheses created by the WCC pipeline. There is no direct way to measure it, as the data set is not complete and the underlying statistical distributions are not available. As an alternative, a stochastic distribution of detections can be generated and used as an input to the Local Association (LA) and Conflict Resolution (CR) processes described in [Kitov, 2026abc]. One of the best ways to generate a stochastic detection set is to use a stochastic time series instead of actual data in the WCC detection process. The choice of MEs and the template signals at the IMS stations associated with them is crucial for the evaluation of the random detection case. The number of MEs can be limited, but their distribution has to correspond to the zone of aftershock sequence of the J29 event shown in Figure 1. For this feasibility study, twenty MEs were selected by their positions and the number of associated stations. These MEs possess templates with the largest SNR values estimated by the IDC. Red circles in Figure 1 present the set of twenty MEs. It is worth noting that the deepest earthquakes are also presented in this set. Black dots represent the set of twelve MEs for the March 11, 2011, Tohoku earthquakes. They are used to study the side-sensitivity of the IMS array stations to large events far away from the studied region of Kamchatka. The IDC parameters for all 32 MEs are listed in Appendix 1.

From Figure 1, it is clear that the 20 MEs may not perform as good as the best 100 MEs in terms of the number of event hypotheses in the XSEL. However, the performance and general properties of the selected 20 MEs are compatible with all other MEs used in the previous study [Kitov, 2026d] and in the prototype WCC routine processing conducted by the IDC.

***WCC detection with stochastic noise***

A stochastic time series is needed for the evaluation of the statistical significance of the event hypotheses created in the WCC processing. An XSEL obtained with a given set of defining parameters has to be compared with the XSEL obtained in a "random" case. The latter includes a set of detections randomly distributed over the processed interval and having SNRcc values obtained from a random distribution in the range characteristic for the WCC detections generated by a stochastic time series. For the WCC pipeline, the best approach is to generate a stochastic time series and process it as an actual data set in the same study.

Generation of a stochastic time series without serial correlations is a case dependent problem [Press et *al.*, 1986]. The selection of the *ran3* Fortran code to generate random numbers converted into a stochastic time series was dictated by the original WCC pipeline code written in Fortran 77. It guaranteed at least 2,000,000,000 random numbers matching the requirements of the data at the IMS array stations and was sufficiently quick not to increase the total calculation time. This random number generator was also used in [Adushkin et *al.*, 2025] to model stochastic noise added to the actual time series for the initial assessment of the gain in the WCC detection threshold.

The SNRcc frequency distributions obtained from a stochastic time series for stations PETK, PDYAR, MKAR, KURK, and WRA are shown in Figure 2. The stochastic time series was added to the actual data for the July 28, 2025 (jdate=2025209). This was a day of relatively low seismic activity within the studied region, and a day approximately 24 hours prior to the July 29 Kamchatka earthquake. The stochastic waveform was scaled to the maximum amplitude of the actual time series in a given hour for each channel and filter individually. The WCC processing has an hourly rate, with an overlap with the next hour by six minutes to provide a smooth transition to the next interval and avoid the edge effects for the cross-correlation time windows (CCWW) and long-term average amplitude (LTA) calculations. The last fifteen minutes of the processed hour are saved for the next hour, and the total processing interval is 81 minutes long. The SNRcc values are estimated for each tome count within the processed hour according to the WCC detection procedure with the total number of samples equal to the sampling rate multiplies by 3,600 seconds. The estimates for the full day of 24 hours are counted in 0.1-unit bins. The obtained curves are shown for each of the 20 MEs, with lower numbers of MEs at some stations due to the late start of operation or longer periods of upgrade and maintenance.

For station PETK in Figure 2a, all nineteen available MEs show nearly similar behaviour with small deviations. There is a peak near 1.4 followed by a quasi-exponential roll-off down to SNRcc level of 3. Then, some curves demonstrate low-amplitude fluctuations, and all the curves except four do not reach the SNRcc level of 4. There are four curves having SNRcc values above 4.0 but below 5.0. From the REB, station PETK is associated with almost all events in the studied area during its operational period. For the Local Association process, is has the maximum event weight of 1.0, which represents the highest probability of a station being associated with the event hypotheses in the designated area. Station PETK is a major anchor for valid event hypotheses. Figures 2b and 2c illustrate the curves similar to those for station PETK. Stations PDYAR and MKAR are two small-aperture arrays. All other small-aperture IMS arrays not shown here are characterized by similar SNRcc frequency distributions, including the maximum SNRcc level below 5.

The full set of principal Event Definition Criteria (EDC) used in the Local Association process defined in [Kitov, 2026d] is listed in Appendix 2. They include the total event weight determined as the sum of the weights of all associated stations. For the events with three to five associated stations, one of these stations must have a weight above 0.855 for the strict case and 0.8 for the weak case of the LA. Also, this mandatory station must have SNRcc larger than 5.0 and 4.5, respectively. Therefore, the stochastic time series practically never produces appropriate SNRcc values at station PETK to match this minimal SNRcc criterion. PETK is practically excluded from the event creation procedure, except for events with 9 or more associated phases where the SNRcc restrictions are lifted.

The EDC are used in the WCC pipeline, and the results of this processing depend on the version of the defining parameters. Since these parameters are the same as in [Kitov, 2026d], all results are directly compatible. Any difference between the original results and those obtained in this study can be considered related to the new parameters such as *StN* and the detection thresholds of 100 and 1000, or to the newly simulated stochastic waveforms. The sets of EDC were estimated in [Kitov, 2026d] from the SNRcc frequency distributions at the IMS arrays using quiet days without REB events and very low numbers of XSEL events. In some cases, these curves are similar to those at PETK.

The mid-aperture array KURK reveals a feature potentially related to the WCC detection algorithm. The curves at KURK have a peak slightly above the detection threshold of 3.3. This is similar to the pattern observed for the detections of the sought signals in the actual data. However, the deep troughs at 3.2 are a new feature for the SNRcc curves. For the actual detections, no troughs are observed. In any case, the SNRcc at KURK are all below the 4.5 level, except for a few cases for IDC orid=21227220.

Figure 2e displays the SNRcc curves for the mid-aperture station WRA. It also serves as an illustration of troughs before the peaks above the detection threshold of 3.3. These troughs are much deeper than at KURK and reach a level of 1 for one of the MEs. All mid-aperture IMS arrays, except for ILAR, demonstrate this feature and no small-aperture arrays show any significant troughs in the SNRcc curves before the detections. The peaks in the SNRcc curves are related to the detection algorithm freezing the LTA when the SNRcc reaches the threshold value. For a stochastic waveform, the SNRcc values calculated without above-the-threshold procedure or with extremely high threshold can be higher of lower than the detection level. The frozen LTA may increase or decrease them relative to their stochastic levels. The effect would be the troughs and the peaks. Figure 3 demonstrates the effect of the detection threshold set at 100 at station CMAR. The peaks and troughs observed for the routine detection threshold of 3.2 disappear. The curves for the routine processing are extended by the detection algorithm to the level above 5.0 for three MEs. The other MEs are limited by the maximum SNRcc of 5. Station CMAR has a relatively low weight and cannon serve as a mandatory contributing station.

The evolution of the SNRcc frequency distribution with *StN* and the detection threshold is shown in Figure 4 for the large-aperture station NOA. This is the largest IMS array designed for far-teleseismic signals. It fails to process regional and near-teleseismic signals using all 42 channels. For regional distances, only one out of seven groups of 6 sensors are used. For teleseismic signals, there are source-specific arrival time corrections to be used instead of theoretical time delays in order to avoid beam-loss. Otherwise, the signals cannot be stacked synchronously. The WCC has no such problem, as the cross-correlation at an individual channel does not depend on the other channels and channel staking has zero time-delay.

Figure 4a shows the setting with the *StN*=100 and the routine detection threshold 3.2. The troughs and peaks are as well developed as those at the mid-aperture IMS arrays. Figure 4b presents the Figure 4a case but with *StN*=1000. The peaks and troughs are almost the same, as is the overall shape. The absence of any difference between the cases with *StN*=100 and *StN*=1000 confirms the consistency of the assumption that the waveform is stochastic. Figure 4c presents the case with *StN*=1000 and a threshold of 100. The corresponding SNRcc curves have no peaks and troughs, as the threshold is so high that no SNRcc value can reach it. These SNRcc curves are truly stochastic, but a comparison with the original XSEL with *StN*=0 requires retaining identical thresholds.

a) Station PETK

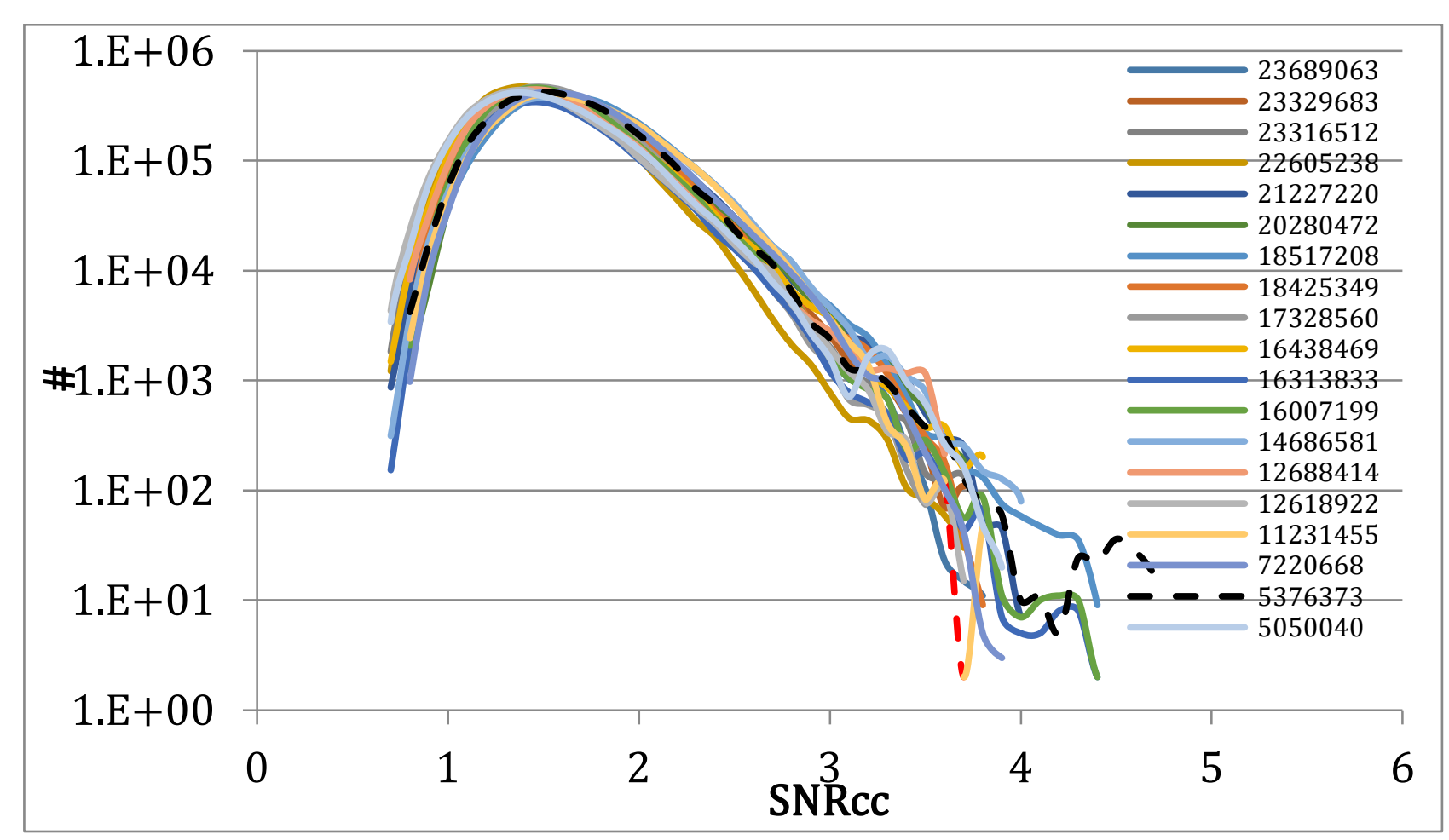

b) PDYAR

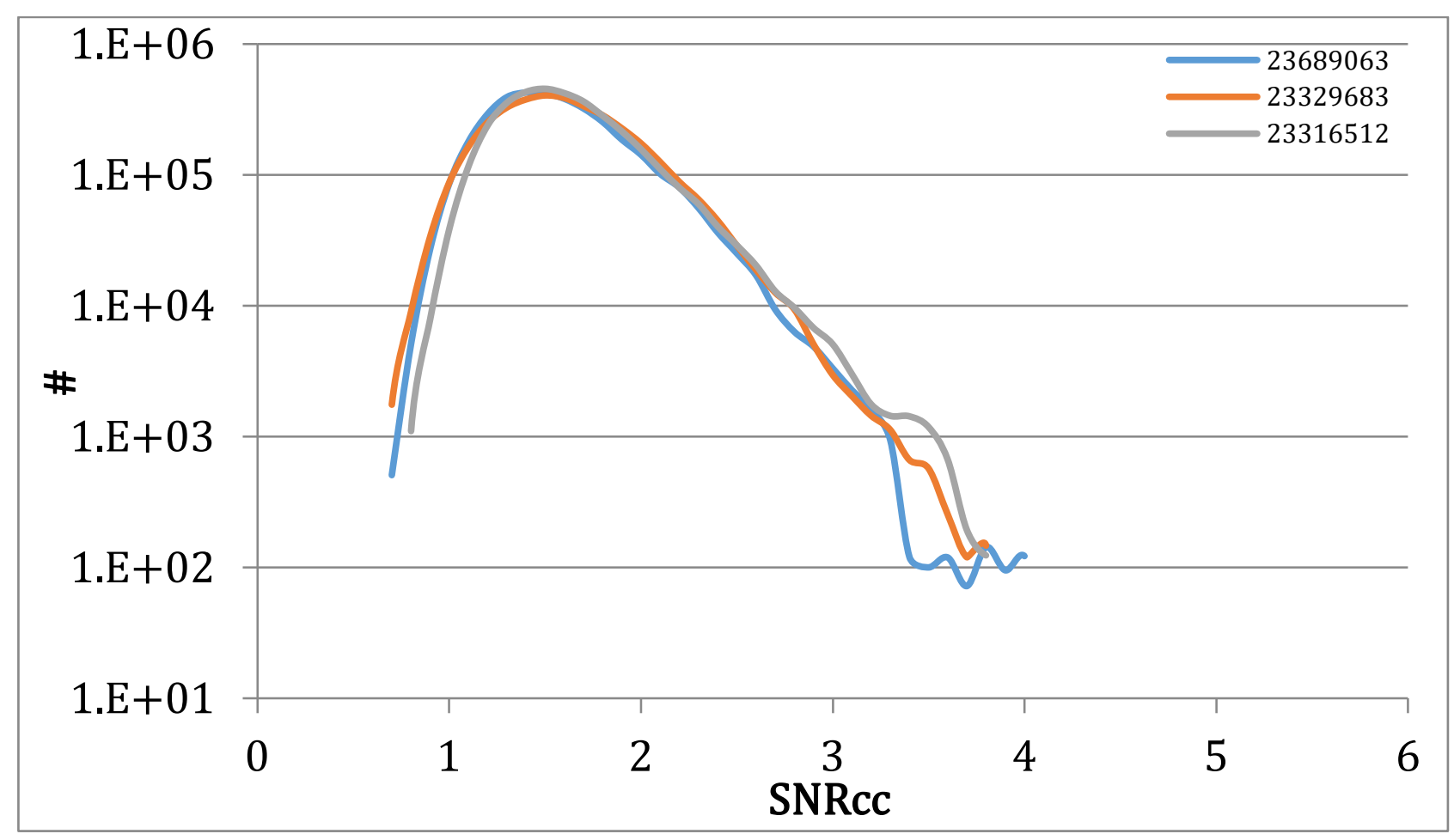

c) Station MKAR

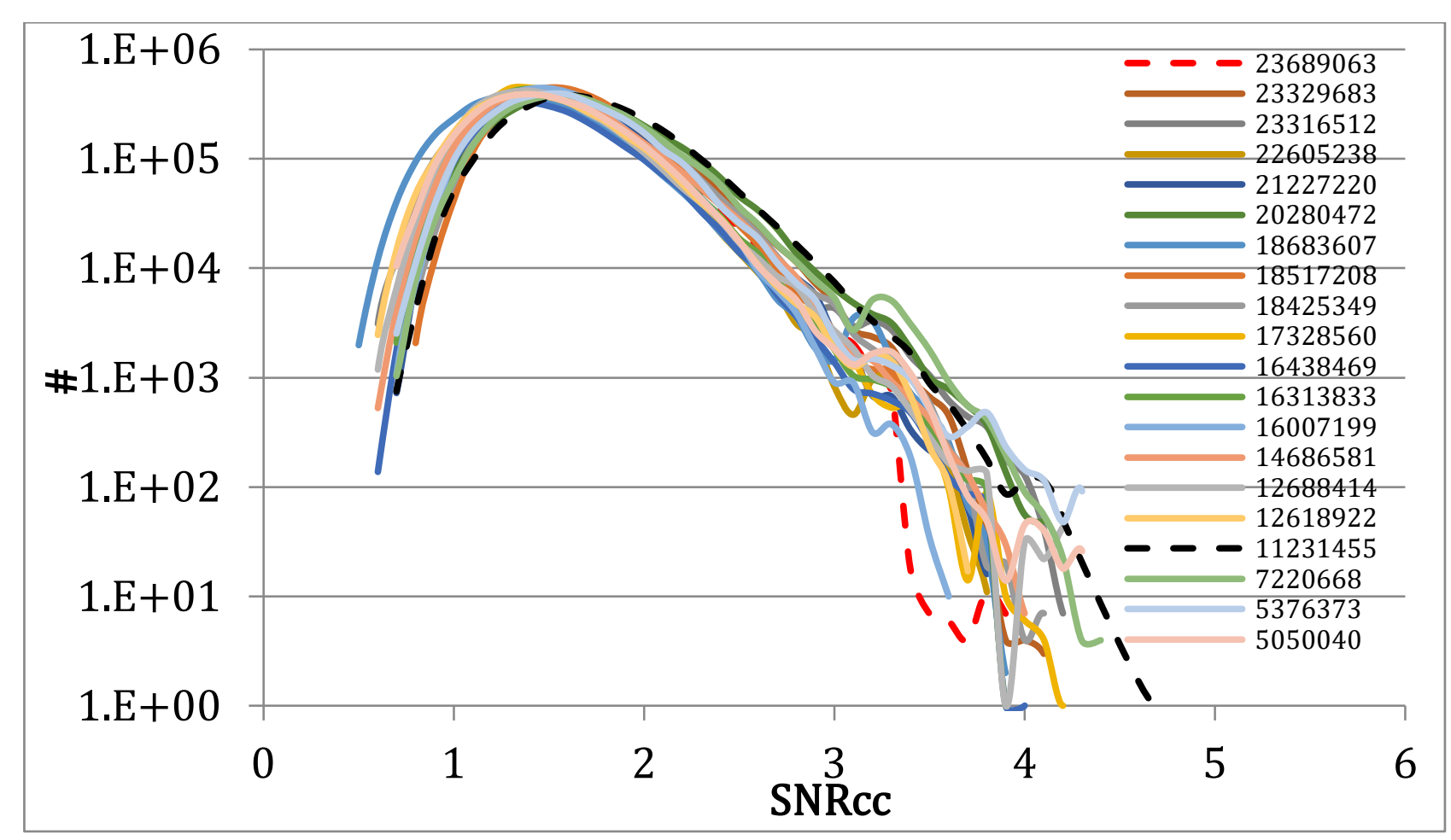

d) Station KURK

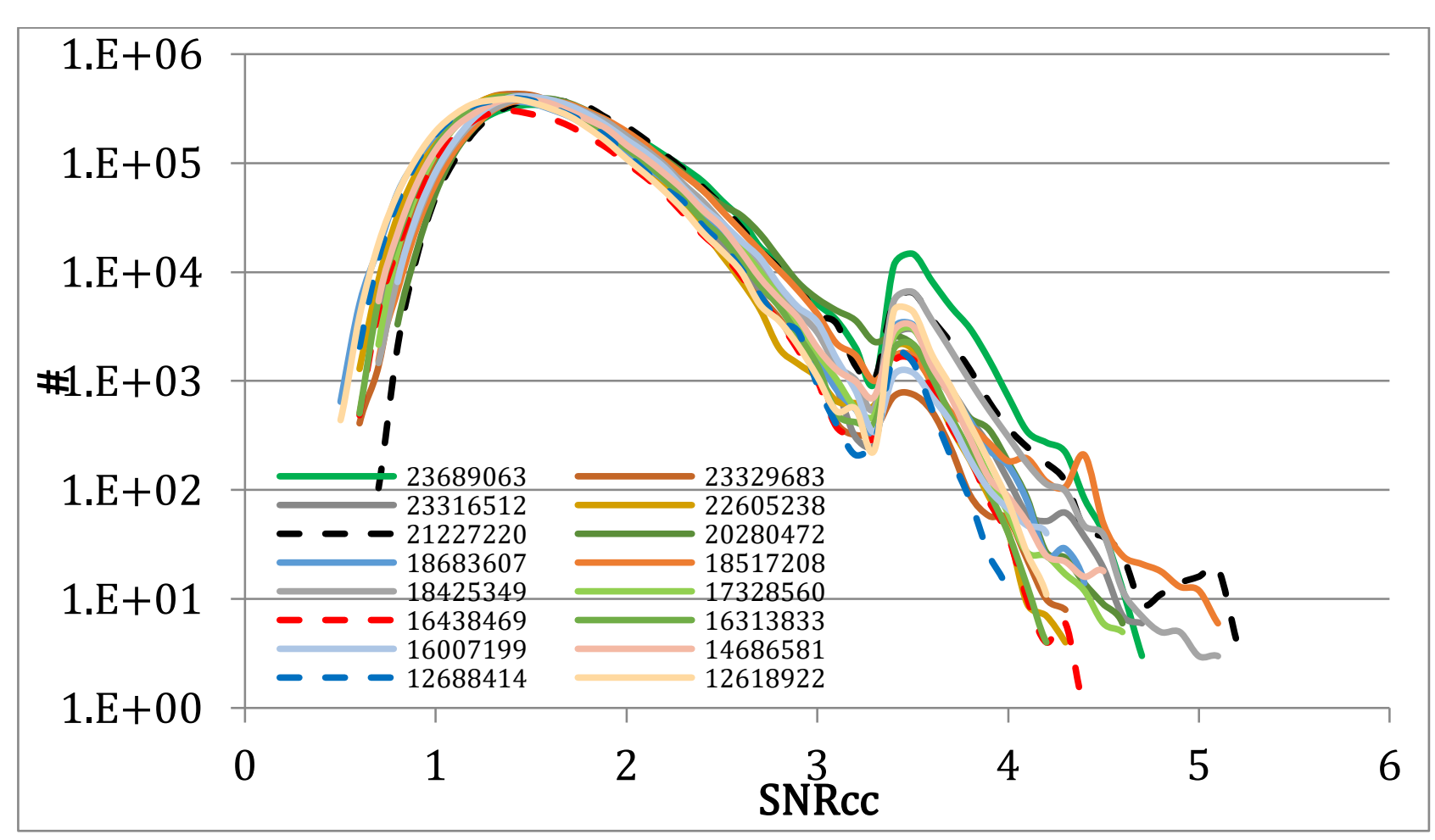

e) Station WRA

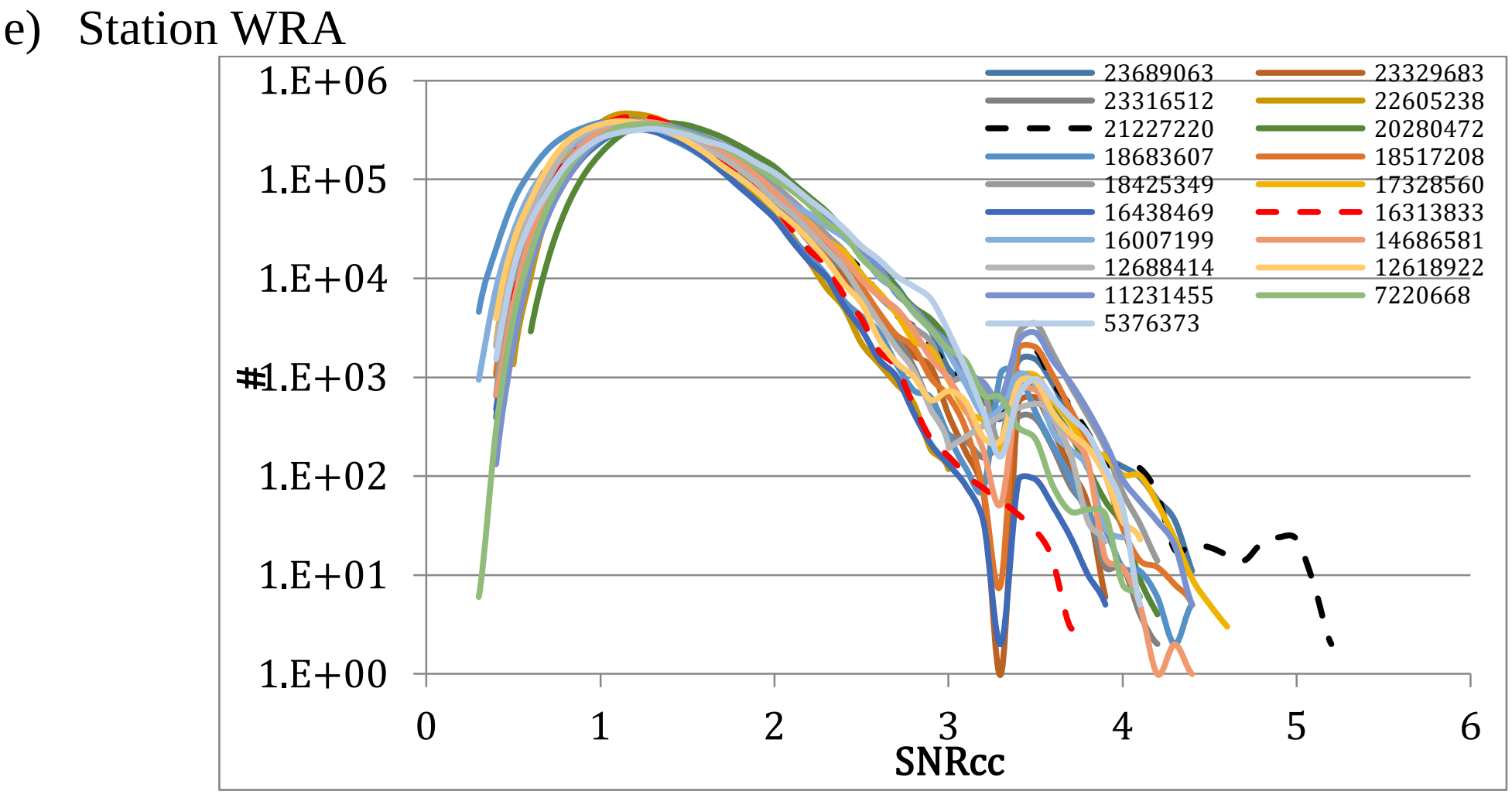

Figure 2. Frequency distribution of SNRcc values at four stations (PETK, PDAR, MKAR, KURK, and WRA) on July 28, 2025. Twenty master events are used. Station may miss some of the MEs due to maintenance works, upgrade, or late start of operation.

a)

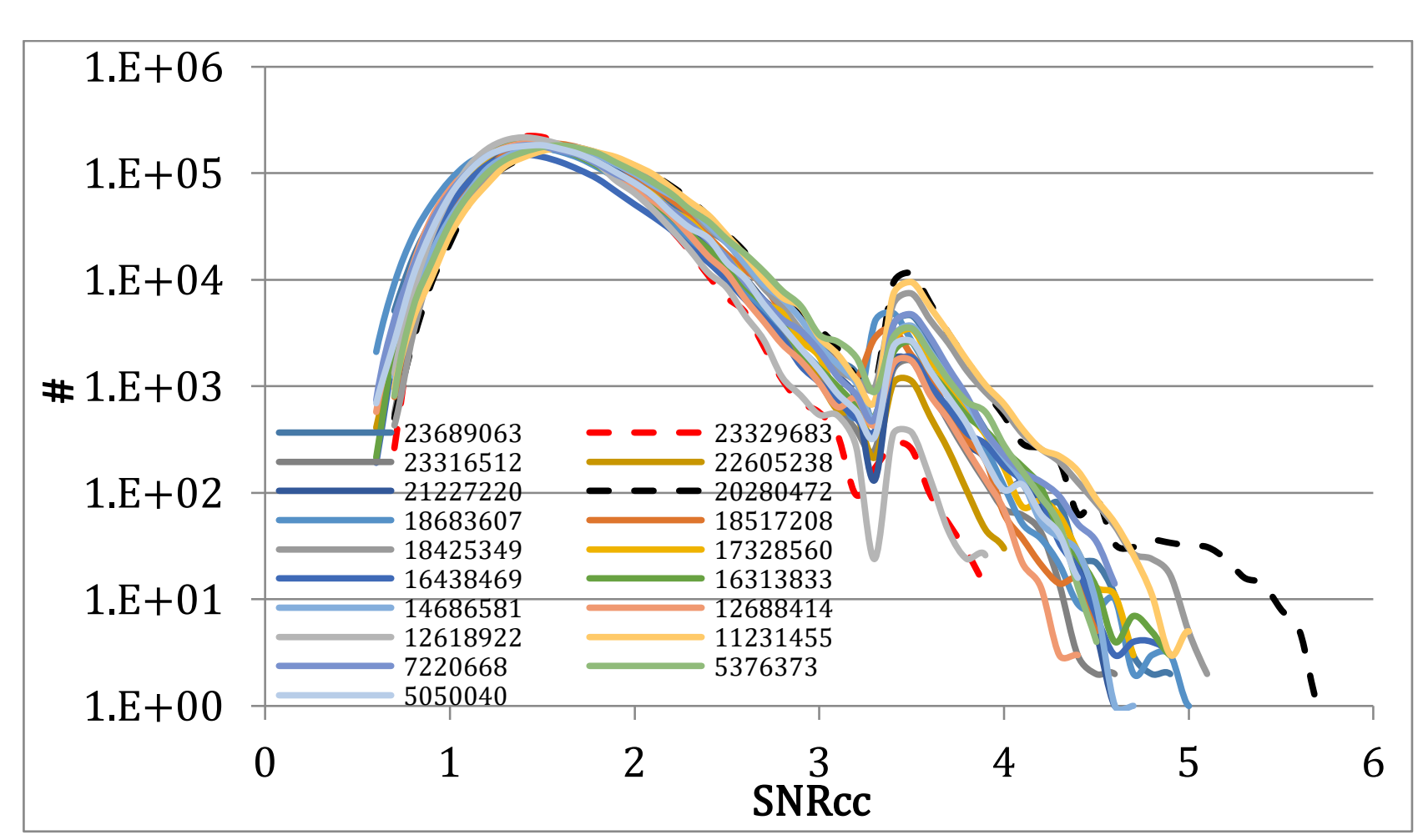

b)

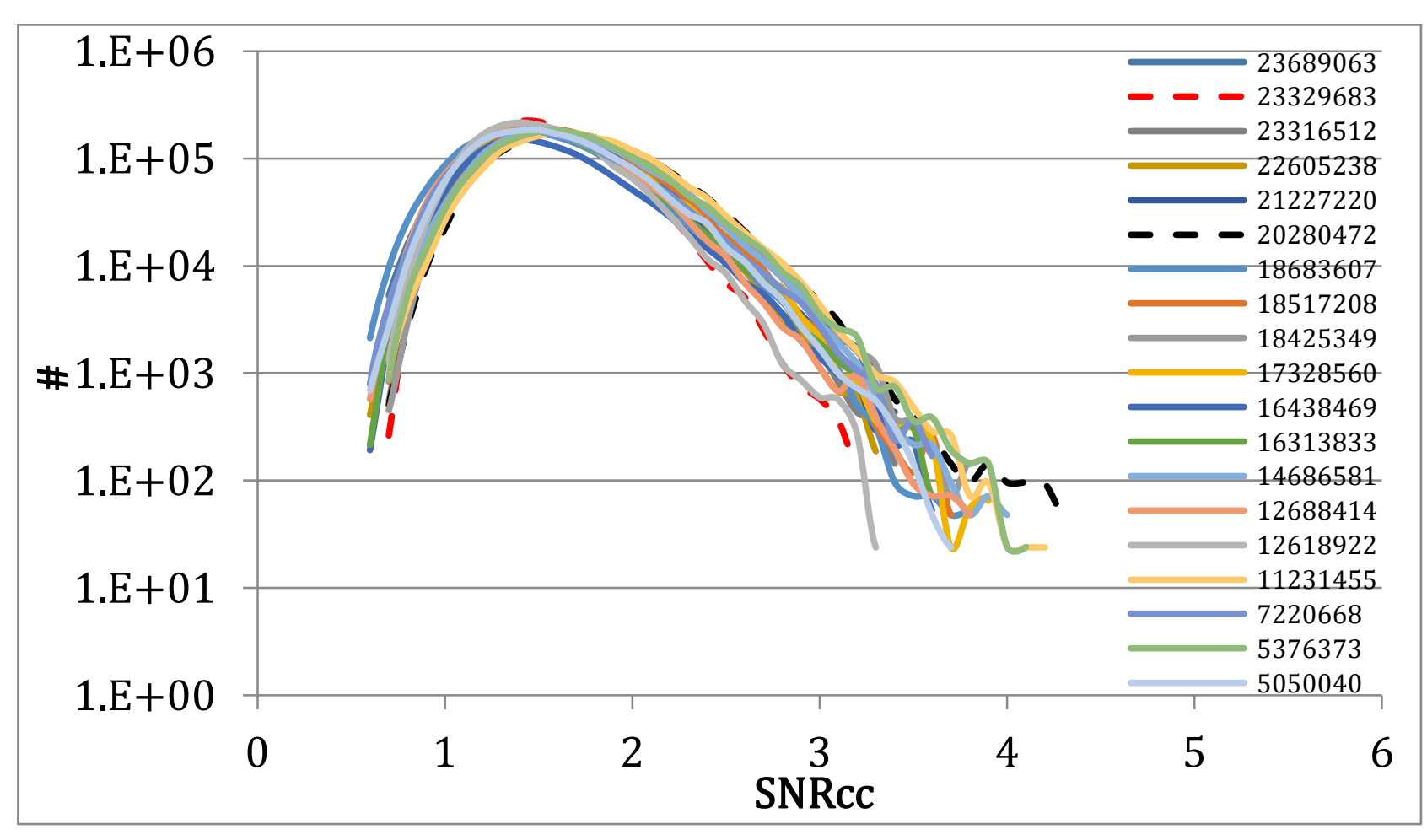

Figure 3. Comparison of the SNRcc frequency distributions with detection threshold 3.1 (a) and 100 (b). Stations CMAR. The effect of threshold and detection procedure is significant.

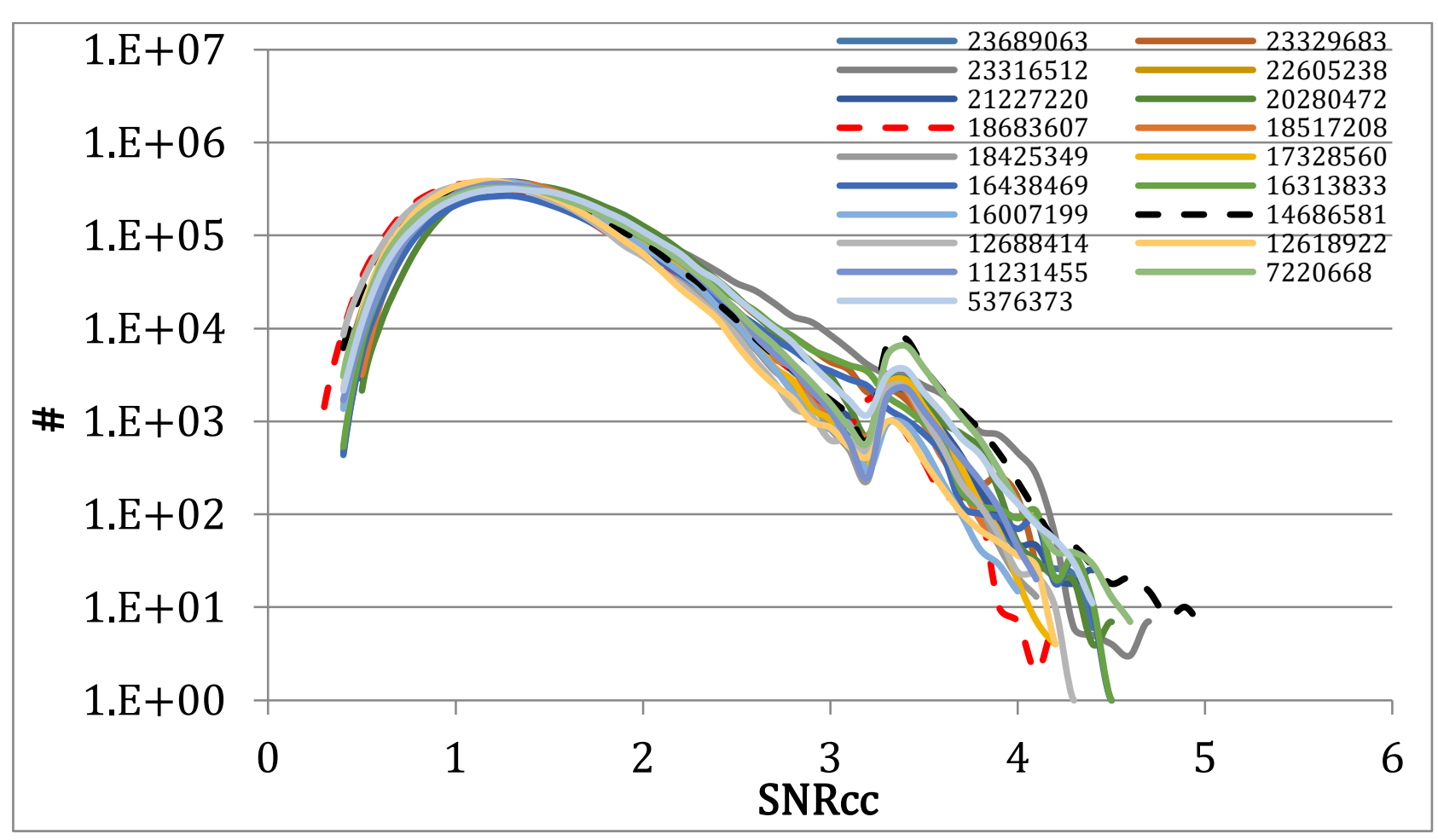

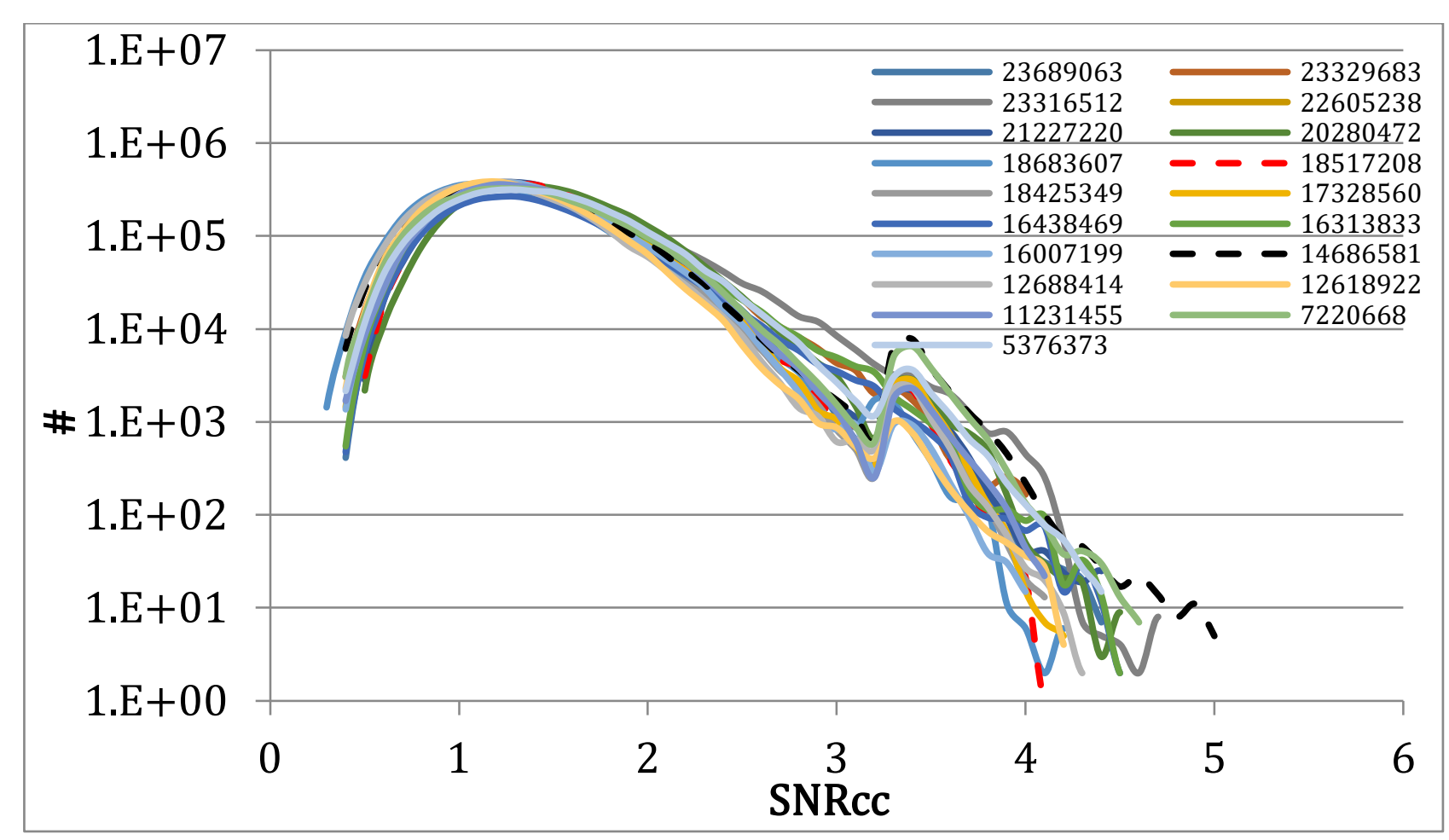

c)

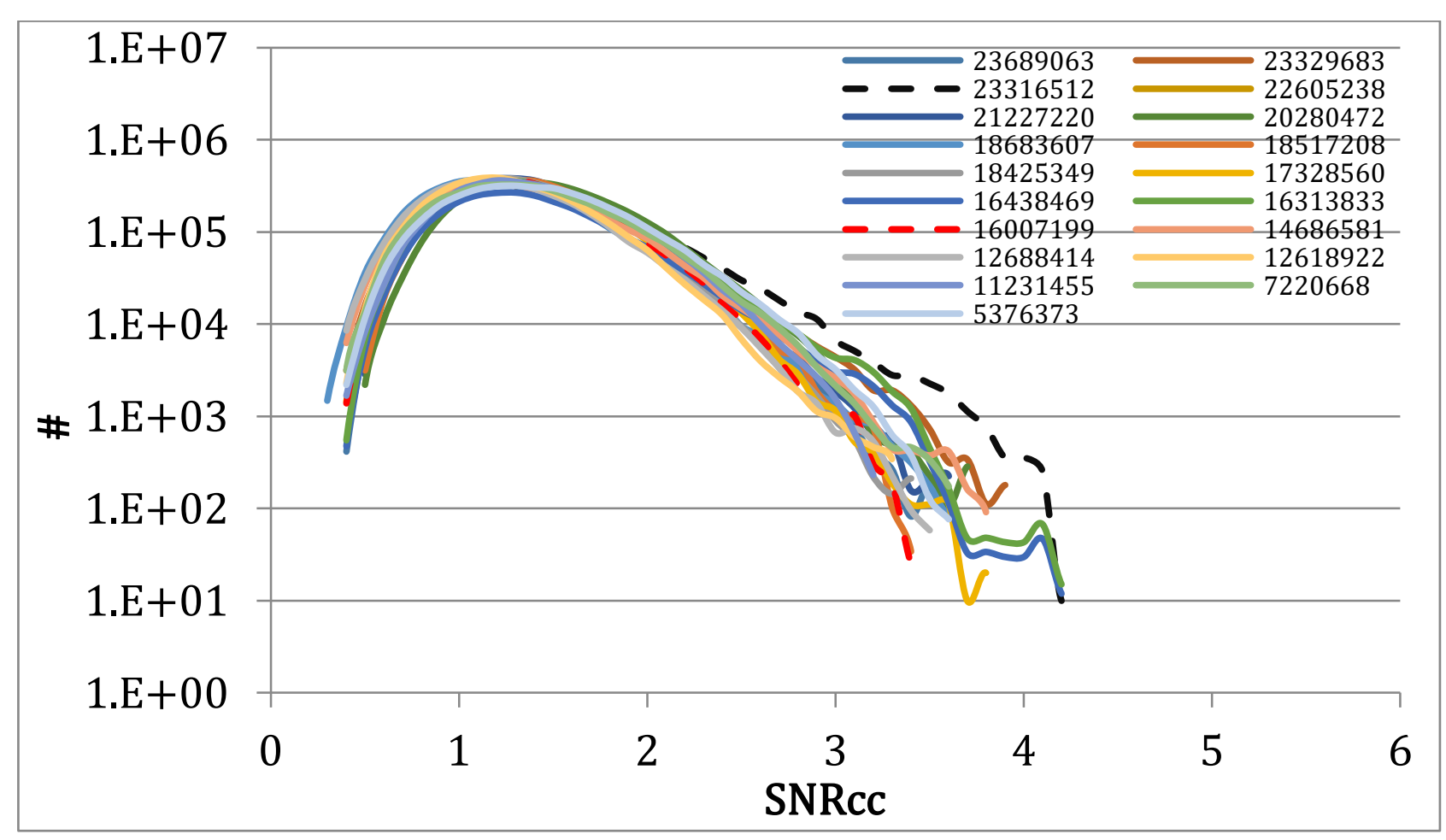

Figure 4. Frequency distributions of SNRcc at station NOA. Three cases: a) *StN*=100, detection threshold 3.2. b) *StN*=1000, detection threshold 3.2. c) *StN*=1000, detection threshold 100.

The principal features of the SNRcc curves generated using stochastic waveforms and shown in Figures 2 through 4 for further phase association are related to the EDC adopted for valid events. Hypotheses with the number of associated stations between 3 and 5 must include a highly reliable station (weight above 0.855 or 0.80) with the SNRcc value above 5.0 (4.5). There is also a requirement for the sum of the SNRcc values of all associated stations. For a three-station event, it must be above 15 (14), and the sum increases by 3.5 for with each additional station.

The rule of a minimal SNRcc of 5 (4.5) at one out of a few best stations is applied to the 3- to 5-station events. For 6- to 9-station events, the station with the high SNRcc is not restricted to the best set. For 10- or more station events, no minimal SNRcc rule is applied. The SNRcc curves for all small-aperture stations have a maximum SNRcc for the best ME below 5.0, with a majority of the 20 MEs generating SNRcc values below 4.5. The mid– and large-aperture arrays generate SNRcc values below 5.5. They also generate an excess of SNRcc values near the detection threshold used in the WCC processing of actual waveforms.

By design, the EDC are intended to prohibit the creation of random events. The SNRcc curves for quiet days were used to estimate the thresholds in the LA to guarantee very low probability of a random event being created. The performance of these preliminary thresholds is tested in the LA and CR processing.

***WCC Local Association and Conflict Resolution with stochastic noise***

Stochastic waveforms with various scaling factors *StN* were processed at all the IMS stations participating in the study of the July 29, 2025, Kamchatka earthquake. Several days were processed in order to understand potential variations in the corresponding XSEL bulletins depending on the seismic activity reported by the IDC. The subsequent procedures of the WCC pipeline were executed to obtain the XSELs – namely, Local Association and Conflict Resolution. There were 20 MEs in this case (see Figure 1), and the CR procedure plays a significant role in determining the quality of the final XSEL bulletin.

In line with the original study of the Kamchatka 2025 megathrust earthquake [Kitov, 2026d], twelve different LA settings were created to cover various magnitude ranges of the XSEL events. There are two LA versions described in Appendix 2 – "weak" and "strict". Each version has six cases defined by the origin time tolerances: 5.0 s, 3.0 s, 2.0 s, 1.0 s, 0.5 s and 0.25 s. This makes twelve version-case pairs arranged in a matrix sequence. Index 1 belongs to the weak version with a 5.0 s origin time tolerance, and index 12 corresponds to the strict version with a 0.25 s tolerance. The first version-case set is aimed at the detection of the weakest seismic events with magnitudes 2.0 and lower. The twelfth set is designed to detect events near the corner magnitude of the recurrence curve obtained from the REB for the studied area. The task is to estimate the XSELs for the stochastic waveforms for all twelve version-case pairs.

Table 1 presents the results of the WCC processing of three sequential days – July 28, 29, and 30, 2025. These three days include almost two days prior to the July 29, 2025, Kamchatka megathrust earthquake, which occurred at 23:24 UTC, and July 30 – the day of the most intense post-seismic activity. Prior to the mainshock, seismic activity was not high, as one can judge from the number of REB events: 7 (two of them pure REB) on July 28 and 8 (2 pure REB) for the 23 hours between 0:00 and 23:00 UTC on July 29. This represents a slightly elevated daily rate in the tail of the aftershock sequence generated by the July 20, 2025, M7.4 earthquake that occurred in the vicinity of the July 29 earthquake. The first six hours on July 30, 2025, likely represented the period of the highest seismic activity measured by the IDC, with 278 REB events within the studied area.

Table 1. The numbers of XSEL events for days 2025209, 20025210, and 2025211. Twelve LA sets (see Appendix 2) and five *StN* values.

| | 2025209 | | | | | 2025210 (0h-22h) | | | | | 2025211 (0h-5h) | | | | |
|---|---|---|---|---|---|---|---|---|---|---|---|---|---|---|---|
| index | *StN* | | | | | *StN* | | | | | *StN* | | | | |
| | 0 | 1 | 5 | 10 | 100 | 0 | 1 | 5 | 10 | 100 | 0 | 1 | 5 | 10 | 100 |
| 1 | 245 | 68 | 6 | 4 | 3 | 281 | 96 | 11 | 14 | 10 | 282 | 98 | 17 | 7 | 9 |
| 2 | 181 | 54 | 2 | 1 | 1 | 232 | 78 | 4 | 3 | 3 | 298 | 102 | 16 | 4 | 5 |
| 3 | 160 | 53 | 2 | 0 | 0 | 217 | 65 | 3 | 1 | 2 | 305 | 106 | 12 | 4 | 2 |
| 4 | 128 | 45 | 2 | 0 | 0 | 164 | 53 | 2 | 1 | 1 | 294 | 105 | 12 | 2 | 2 |
| 5 | 83 | 32 | 2 | 0 | 0 | 114 | 40 | 2 | 1 | 1 | 279 | 99 | 11 | 2 | 0 |
| 6 | 55 | 27 | 2 | 0 | 0 | 76 | 33 | 2 | 1 | 1 | 251 | 80 | 8 | 0 | 0 |
| 7 | 55 | 20 | 1 | 0 | 0 | 60 | 26 | 1 | 1 | 0 | 148 | 43 | 7 | 0 | 0 |
| 8 | 46 | 19 | 1 | 0 | 0 | 51 | 17 | 0 | 0 | 0 | 146 | 46 | 6 | 0 | 0 |
| 9 | 38 | 18 | 1 | 0 | 0 | 51 | 14 | 0 | 0 | 0 | 146 | 44 | 5 | 0 | 0 |
| 10 | 25 | 14 | 0 | 0 | 0 | 29 | 13 | 0 | 0 | 0 | 142 | 45 | 6 | 0 | 0 |
| 11 | 17 | 7 | 0 | 0 | 0 | 22 | 11 | 0 | 0 | 0 | 124 | 41 | 5 | 0 | 0 |
| 12 | 9 | 5 | 0 | 0 | 0 | 17 | 7 | 0 | 0 | 0 | 111 | 39 | 3 | 0 | 0 |

The results of the WCC processing in Table 1 demonstrate variations in the twelve XSELs depending on the noise scaling factor *StN*. These XSELs are obtained in the LA and CR processes applied to the detection lists generated by random waveforms with the respective *StN* values. In the *StN*=0 case, only the actual data were used. This is a reference case to be compared with the results obtained with a random time series added. For the zero *StN* case, the first pair - the one with the least restrictive EDC and the widest origin time tolerance of 5.0 s - has a total number of XSEL events ranging from 245 on July 28 to 282 for the first six hours of July 30. This first pair's XSEL for the first six hours of July 30 has likely reached the largest possible number of events per an ME for the current WCC setting of the detection, LA, and CR requirements.

This can be the result of the detection thresholds and the rules governing the minimal allowed spacing between subsequent arrivals in the WCC detection process. The hourly detection rate is bounded from above by 120 and from below by 30 per hour. Such rates are sufficient for the creation of around 50 events per hour or 300 per six hours. A similar saturation limit is likely reached in the IDC automatic processing based on its EDC and processing requirements. A total of 196 out of 278 REB events have seed events in the automatic SEL3 bulletin, and 82 REB events are added by IDC analysts. There are many more actual events occurring during this time period, but they cannot be detected due to the high amplitude seismic noise and the limit on the number of detections at a station.

For the 20 MEs, this number can be larger, but the CR process effectively eliminates the multiple solutions for the same physical events and the aftershock zone actually belongs to a few out of the 20 MEs. This effect can be illustrated by the numbers of events in the final XSEL for individual MEs compared to the numbers generated in the LA process before the CR is applied listed in Table 2. The ME with the IDC orid=23689063 generated 444 event hypotheses, the highest number among all the 20MEs, and 73 of them were promoted to the final XSEL. The second best ME generated 389 event hypotheses in the LA and 42 appeared in the final XSEL. The other 18 MEs were not so effective for various reasons from a geographical position relative to the most intense part of the aftershock sequence to the lower numbers of the associated templates. The ME with the IDC orid=21227220 generated 177 hypotheses after the LA and no events were promoted to the XSEL. The ME with the IDC orid=22605238 generated no hypotheses in the LA although it was a good ME during the pre-seismic period. The extraordinarily high ambient noise consisting of signals coherent with the templates can be the reason for such behaviour. Only the closest station PETK generated a detection list for this ME.

Table 2. Comparison of the individual XSELs for the 20MEs obtained in the LA process and the final XSEL after the CR. *StN*=0.

| orid | **23689063** | **23329683** | **23316512** | **22605238** | **21227220** | **20280472** | **18683607** | **18517208** | **18425349** | **17328560** | **16438469** | **16313833** | **16007199** | **14686581** | **12688414** | **12618922** | **11231455** | **7220668** | **5376373** | **5050040** |
|---|---|---|---|---|---|---|---|---|---|---|---|---|---|---|---|---|---|---|---|---|
| LA | **444** | **335** | **401** | **0** | **177** | **300** | **270** | **226** | **292** | **298** | **327** | **393** | **361** | **211** | **229** | **389** | **66** | **268** | **107** | **306** |
| CR | **73** | **6** | **16** | **0** | **0** | **5** | **4** | **6** | **4** | **14** | **15** | **17** | **19** | **4** | **12** | **42** | **8** | **16** | **6** | **14** |

The decrease in the number of XSEL events for all the LA version-case from index 1 to index 12 is well illustrated in Table 1. The increase in *StN* from 0 to 1 leads to a dramatic two- to four-fold decline in the XSEL numbers. The effect of the random noise with the maximum amplitude equal to the maximum amplitude of the one-hour interval of the actual waveform in a given channel-filter configuration demonstrates that it likely contains no components coherent to the templates. In the case of a high noise and template coherence, the effect of the equal amplitudes is devastating to the XSEL [Kitov, 2026c].

For the *StN*=5, the XSEL numbers drop to the level below 5% to 10% of those of the *StN*=0. For the pre-seismic period, the XSELs for the indices from index 10 to index 12 are characterized by the absence of event hypotheses. For the post-seismic period, the numbers are above zero for all the indices. This has to be the effect of the signals from the large-magnitude aftershocks coherent with the template in the high-amplitude random noise. The matched filter detector is based on this assumption. It can find similar signals deep in the random noise. The results for *StN*=10 and 100 are very similar and demonstrate the absence of XSEL events in all tolerance cases of the strict LA version. For the weak LA version, there are a few XSEL events generated with their number decreasing with the origin time tolerance. They are likely related to the random and thus noise-amplitude independent generation of detections and event hypotheses. They do not depend on the templates but can fluctuate with the random noise realizations.

The XSELs for the strict LA version always have zero random events for *StN*=100. The statistical significance of the XSEL events for this version is very high as the random generation is prohibited. For the weak LA version, there is a probability of a random event being created, but it is less than 5% and does not affect the uncertainty of the precursory parameters used in [Kitov, 2026d]. The source of false event hypotheses is not only random noise but also the side-sensitivity of the WCC detector at array stations to the high-amplitude signals from the events not related to the studied area. Table 2 confirms the potentially significant input from such sources with the MEs far from the aftershock zone of the 2025 Kamchatka mega-earthquake. They create event hypotheses close to their own positions but far away from their real physical hypocenters. This effect deserves a special investigation to evaluate its influence on the results in [Kitov, 2026d].

Overall, the WCC detector inherently generates false detections from the stochastic waveforms. The distribution of the SNRcc values in Figure 4c, obtained in the case of *StN*=1000 and the detection threshold of 100, demonstrates that the quasi-exponential roll-off ends at the maximum values of approximately 4.0. The LA requirement of the SNRcc to be above 5.0 for the strict and above 4.5 for the weak version would prohibit the creation of any XSEL event hypotheses. The real thresholds are much lower and the WCC detection procedure with a frozen LTA extends the SNRcc distribution to the level of 5.0 for the stochastic time series. There can be an extremely rare chance when a specific realization of stochastic time series can produce an event hypothesis with the number of associated stations less than 10. Otherwise, all generated hypotheses must have 10 or more associated stations, when the requirement of the minimal SNRcc of 5.0 or 4.5 is lifted. The probability of such events for a given LA version directly depends on the origin time tolerance. As a result, no XSEL event hypotheses are produced for the stochastic waveforms generated with *StN*>10 by the strict LA version with the tolerances below 3.0 s. For the weak version, there are insignificant numbers of XSEL events generated for the tolerances below 2.0 s. The results reported in [Kitov, 2026bd] are not biased by the random false detections.

### *Distance-dependent efficiency of MEs*

Table 2 illustrates the relatively lower efficiency of the MEs outside the aftershock area. The distance between an ME and a sought event is crucial for the shape similarity of the template and the sought signals at array stations. The template-sought signal cross-correlation coefficient (*CC*) decreases with distance relative to that for the ambient noise [Arrowsmith and Eisner, 2006; Baisch *et al*., 2008]. Consequently, the WCC detection probability decays with distance, *ceteris paribus*. However, the increase in the standard SNR of the sought signal, *i.e*., the increase in the sought event magnitude, can compensate for the distance–dependent *CC* decrease. This is the WCC detector's side-sensitivity when the *CC* increases at a few random channels in sync with that of the template and the sought signal *CC*, with the others retaining low *CC* values. As a result, the average *CC* trace generates false WCC detections at many stations. The differential arrival times at the stations associated with a given ME deviate progressively more with the ME-sought event distance, and the LA may use these false detections to produce false event hypotheses far away from the physical positions of the sough events.

Therefore, the MEs outside the zone of the most intense post-seismic activity can generate many event hypotheses in the LA process as their templates have lower but still sufficient similarity with the signals from this seismically active area. These hypotheses have lower statistical significance and lose in the CR to the hypotheses within or closer to the area. However, these remote MEs can generate winning valid events in the zones of their responsibility. There exists a distance range where the CR process has to select between two MEs with very close defining parameters such as SNRcc, which are intrinsically prone to measurement uncertainty. For such hypotheses in the gray zone, the CR process may result in a wrong choice. Consequently, the XSEL may lose a valid event matching the REB without any change in the total number of events. As a result, the REB match statistics before the CR is also a useful parameter to evaluate the WCC overall performance. Such erroneous CR decisions can also be mitigated by the addition of new MEs in the gray zone. This requires more computer power or longer processing time. There are around 1000 MEs in the studied area of the 2025 Kamchatka earthquake in the prototype routine WCC processing at the IDC.

For a given *StN* value, the number of XSEL events decreases from index 1 to index 12. This is the main feature of the twelve version-case pairs. They define the increasing lower magnitude boundaries for the corresponding XSELs leading to a decrease in the number of XSEL event hypotheses. For a short period of a few days, the dependence of the number of events across the twelve XSELs can be locally biased by the underrepresentation of events in the corresponding samples. In the original study [Kitov, 2026d], for longer periods of weeks or more, these curves demonstrate an exponential decay with coefficients of determination up to 0.99, and thus, the version-case index can be converted into a linear magnitude scale. Figure 5 presents the number of XSEL events as a function of the index for *StN*=0. Overall, the exponential regression lines for the pre-seismic period are characterized by high coefficients of determination. The post-seismic period immediately after the mainshock is characterized by a step between two versions and almost constant XSEL numbers within the same version [Kitov, 2026bd].

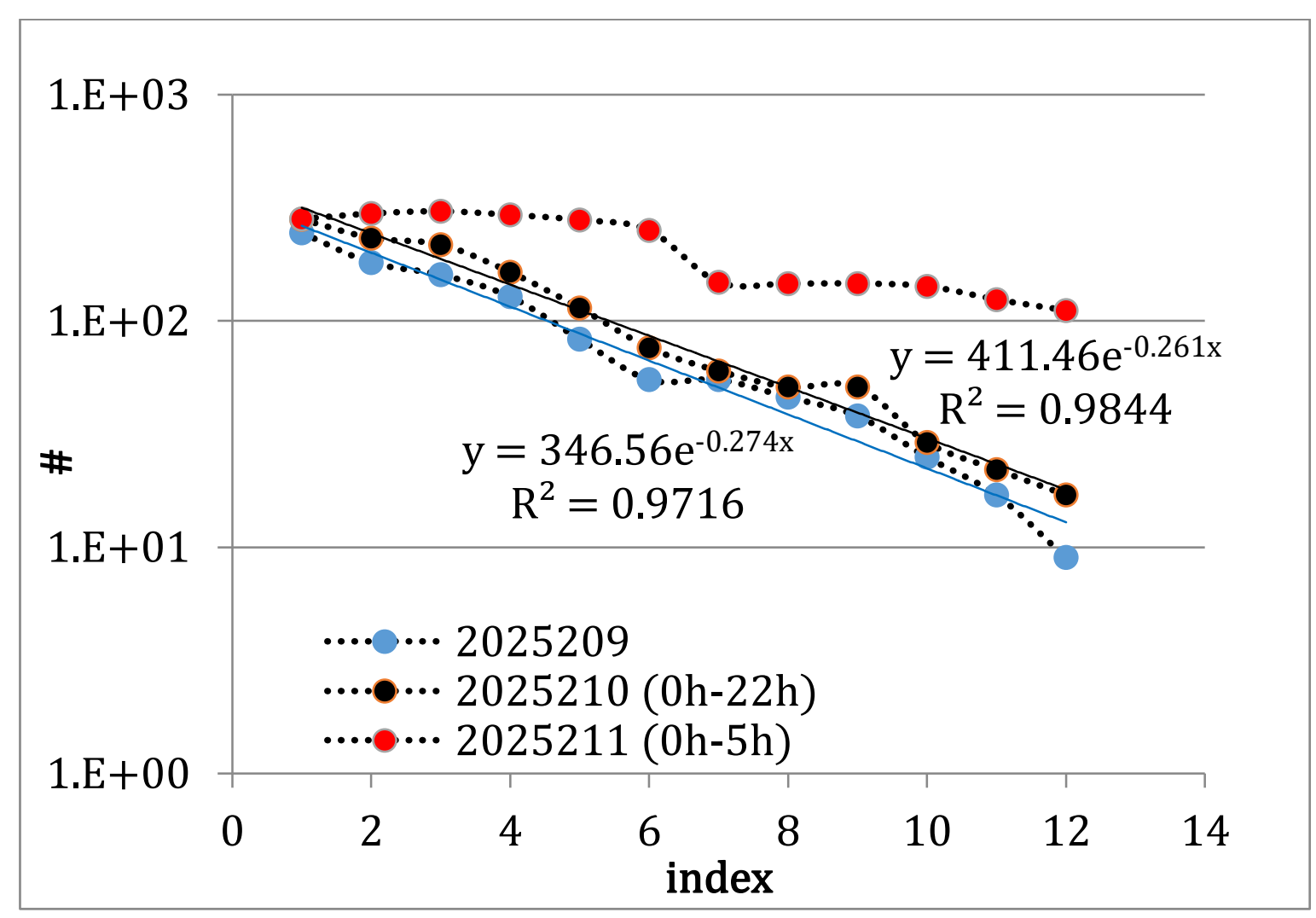

Figure 5. Number of XSEL events as a function of the index of the version-case pair. *StN*=0.

Such behaviour indicates that for the highest rate of aftershocks in a relatively small area the origin time tolerance window does not affect the XSEL much. For these aftershocks, the actual magnitude boundaries for all tolerance cases of a given LA version are larger than the corner magnitude of the respective recurrence curve. For the WCC processing, the XSEL magnitude boundaries are defined by this actual magnitude threshold and the smaller events are almost fully suppressed. For a given version, all XSELs are approximately equal. The difference between the two versions is expressed in the difference in the share of actual events matching the EDC for the WCC.

### *The XSEL statistics for the REB-matching events and the number of new XSEL events*

The results of the random generation of detections and XSEL event hypotheses in the WCC processing have to be reflected in the statistics of matched REB events. The latter are generated in the interactive review by IDC analysts. The term “match” means the fact that the XSEL and REB detections are within 20 s of each other. This rule follows from the retiming interval allowed by the IDC rules for an automatic detection without the creation of a new detection. This value follows from the statistics of arrival time difference between the SEL3 and REB for the same physical signals when the former were also associated with the REB events [Saragiotis and Kitov, 2020]. The station-to-station XSEL-REB comparison is a natural operation since they both use the same IMS stations but different detection lists.

For an REB event to be matched by an XSEL event, one REB-matched P-phase is sufficient, as adopted by the IDC for the comparison of the automatic SEL3 bulletin and the REB. The XSEL uses only P-phases for the cross-correlation calculations, but the length of templates can be sufficient to include many secondary phases. The template length is an important parameter because the difference in the group and phase velocities between various secondary phases results in a rapid change in the wavefield. The change in the shape of the evolving sought signal puts stringent restrictions on the relative position of the respective sought event. The level of cross-correlation with the template drops as rapidly as the sough signal changes its shape.

It is important to stress that the matching XSEL and the matched REB events do not have to be close in space, as the IDC has a strict rule to fix the events with a high uncertainty of depth estimates to the surface. Consequently, a significant bias up to 10° can be introduced in the respective IDC hypocenter locations. The locations of the XSEL events are close to their respective MEs. False XSEL events can also match REB events. Therefore, the number of false XSEL events has to be minimized to the greatest extent possible. The match statistics are important for the demonstration of statistical power of the XSEL events and the reliability of the WCC processing.

The statistics of new XSEL events, which are not matching any of REB events, are important for the evaluation of the random events input. The number of random events has to change with the origin time tolerance case for a given version, as Table 1 illustrates. The total number of new XSEL events has to decrease with the tolerance case when the magnitude boundary is not elevated by high seismic activity. In that case, the number of new XSEL events does not depend on the tolerance and differs between versions.

Table 3 presents extensive statistics of the XSELs for *StN*=0 as obtained during the period between July 28 and July 30, 2025. There are two relatively quiet days of July 28 and 29 (23 hours). July 30 is a day of the highest seismic activity. The last hour of July 29 is not processed. The quiet days are processed as one interval. July 30 is split into four 6h intervals with a decaying activity – from a total of 278 REB events in the first interval to 160 events in the last one. Two configurations with of 20 MEs and 100 MEs are used. The latter is the set of the best 100 MEs used in [Kitov, 2026d], as shown in Figure 1. This allows for the evaluation of the effect of the MEs density on the statistics.

Column "REB/SEL" in Table 3 displays the total number of events and the number of events built starting from the SEL3 seed events. The difference between these two numbers is the number of pure REB events built by IDC analysts. This is a measure of the IDC automatic processing efficiency. For example, on July 30 the ratio of the number of the pure REB to the SEL3 seeds is 0.42, 0.26, 0.27, and 0.33 for the four subsequent 6h periods.

The XSEL match statistics are shown for all twelve version-case indices. The first number in the cells is the number of matched REB events, the second number corresponds to the matched pure REB events, and the third number is the new XSEL events. For the two quiet days, the match rate for all versions-cases-configurations is from 70% (5 out of 7) 100%. The total number of XSEL events and the number of new events decrease with the version-case index. The difference between the 100 MEs and 20 MEs configurations is from twofold for the weak version to fivefold (18/4) for index 12. This is important for the statistical reliability of the ratios of the number of XSEL events between the weak and strict versions for the same origin time tolerance used in [Kitov, 2026bd] as a precursory variable of the 2013 Sea of Okhotsk and 2025 Kamchatka mega-earthquakes. The configuration of 20 MEs is too sparse to be used for these purposes. The weakest XSEL events have to be much closer to the MEs which can detect them.

Table 3. Statistics of the REB matches and new XSEL events during the period between July 28 and July 30, 2025.

| Day | #ME | Hours | REB/SEL | Match REB/pure REB/NEW | | | | | | | | | | | |
|---|---|---|---|---|---|---|---|---|---|---|---|---|---|---|---|
| | | | | 1 | 2 | 3 | 4 | 5 | 6 | 7 | 8 | 9 | 10 | 11 | 12 |
| 2025209 | 20 | 0 to 23 | 7/5 | 5/0/239 | 6/1/172 | 6/1/151 | 6/1/119 | 6/1/71 | 5/0/48 | 5/0/47 | 5/0/37 | 5/0/31 | 5/0/17 | 5/0/11 | 5/0/4 |
| 2025209 | 100 | 0 to 23 | 7/5 | 7/2/457 | 7/2/368 | 6/1/306 | 7/2/225 | 7/2/147 | 6/1/91 | 7/2/116 | 7/2/81 | 6/1/69 | 5/0/46 | 5/0/28 | 5/0/18 |
| 2025210 | 20 | 0 to 22 | 8/0 | 8/0/265 | 8/0/215 | 8/0/201 | 8/0/147 | 7/0/101 | 6/0/65 | 6/0/48 | 6/0/35 | 6/0/37 | 6/0/19 | 6/0/12 | 6/0/6 |
| 2025210 | 100 | 0 to 22 | 8/0 | 8/0/475 | 8/0/392 | 8/0/337 | 8/0/260 | 8/0/182 | 8/0/126 | 8/0/108 | 8/0/76 | 8/0/62 | 7/0/40 | 7/0/20 | 7/0/22 |
| 2025211 | 20 | 0 to 5 | 278/196 | 204/33/70 | 210/39/75 | 206/36/82 | 193/32/80 | 187/32/68 | 180/25/46 | 163/18/5 | 154/14/9 | 152/12/7 | 149/12/4 | 132/12/4 | 123/12/3 |
| 2025211 | 100 | 0 to 5 | 278/196 | 242/55/85 | 231/51/110 | 228/50/107 | 232/49/97 | 221/36/83 | 209/37/66 | 196/30/9 | 193/25/11 | 186/26/10 | 181/27/6 | 173/25/4 | 162/21/4 |
| 2025211 | 20 | 6 to 11 | 188/149 | 158/21/79 | 161/24/73 | 161/23/76 | 160/23/67 | 160/22/63 | 150/20/48 | 134/11/8 | 124/10/7 | 125/12/8 | 117/9/7 | 117/11/6 | 101/10/5 |
| 2025211 | 100 | 6 to 11 | 188/149 | 171/29/115 | 177/31/122 | 176/31/116 | 171/23/109 | 170/29/75 | 171/28/60 | 151/18/16 | 153/17/13 | 149/16/13 | 143/16/12 | 143/20/8 | 132/14/6 |
| 2025211 | 20 | 12 to 17 | 175/138 | 149/22/97 | 151/19/87 | 150/17/92 | 147/14/71 | 147/16/57 | 145/15/47 | 141/10/5 | 133/10/6 | 129/10/6 | 123/8/4 | 109/4/3 | 104/5/1 |
| 2025211 | 100 | 12 to 17 | 175/138 | 164/27/128 | 169/31/117 | 165/28/105 | 164/27/93 | 163/26/81 | 159/22/62 | 154/17/14 | 151/18/12 | 150/17/14 | 146/15/7 | 140/15/8 | 136/10/4 |
| 2025211 | 20 | 18 to 23 | 160/120 | 125/18/98 | 125/18/87 | 122/15/94 | 130/16/75 | 124/15/60 | 114/14/48 | 115/12/9 | 113/9/7 | 111/12/6 | 105/7/4 | 98/7/2 | 87/5/3 |
| 2025211 | 100 | 18 to 23 | 160/120 | 148/28/111 | 143/30/108 | 140/26/111 | 145/24/103 | 143/27/89 | 139/21/67 | 133/20/13 | 134/18/14 | 133/15/10 | 129/15/7 | 122/9/5 | 117/9/5 |

For July 30, 2025, the number of the matched REB events depends on the MEs configuration, LA version, and origin time tolerance. For the first 6h interval, the match result for index 1 is not the highest for the 20 MEs configuration. This is the consequence of the largest tolerance forcing an increase in the number of phases for the CR process. Index 2 maximizes both the total number of matched REB events (a total of 204 out of 278) and the number of pure REB events matched by the XSEL (39 out of 82). The former number is larger than the corresponding number for the SEL3 and includes many pure REB events. The WCC processing version for the current study was adjusted to lower detection rates pertaining to the earthquake preparation process. There is a version with reduced spacing between consecutive detections to process the intense aftershock activity, but the focus of this study is to assess the performance of the low-seismicity version.

The respective numbers for the 100 MEs configuration are always higher with the best match rate of 244 out of 278 REB event matched for index 1. The match for pure REB events stays at 55 out of 82. The REB match rate in both categories decreases with the index, but stays high even for index 12 –161 and 21, respectively. These results confirm high statistical power of the XSEL events – most of them are real as the REB match suggests.

The new XSEL events are the only ones to be potentially related to the random detections and related event hypotheses. For the quiet days, their number for the twelve indices follows the exponential roll-off as was shown in [Kitov, 2026d]. It is a prominent feature for the recurrence curves. This exponential decay is also a characteristic of the random detection distribution. The total numbers of the XSELs events are much higher than in the random case listed in Table 1. For the strict version, the number of new XSEL events is always above zero, *i.e*., they are not randomly generated. A very specific feature of July 30 seismic activity is an almost case-independent total number of XSEL events for a given version. The new XSEL events also follow this behaviour, except maybe the shortest tolerance of 0.25 s. For the random case, this number would decrease rapidly with the tolerance decreasing from 5.0 s to 0.25 s.

The statistical estimates in Table 3 do not contradict the assumption that the random generation of XSEL events does not affect both the number of matched REB events and the number of new XSEL events. The effect of random detection is negligible and the statistical significance of the XSEL hypotheses is high enough to be used in the seismological and geophysical research.

**Side-sensitivity of the WCC at array stations**

The effect of high-amplitude noise generated by a mega-earthquake and its aftershocks on the WCC detection process was studied in detail in [Kitov *et al*., 2026; Kitov, 2026c]. This type of noise consisting of high-amplitude regular phases can be an extremely powerful obstacle to WCC detection when it is coherent with the sought signal. For array stations, however, there are situations when such noise becomes quasi-stochastic and improves detections conditions [Kitov, 2026c]. Therefore, it may serve as a quasi-stochastic noise sample when the plane wavefronts of its constituent regular phases are orthogonal to the template and sought signals at an array. The March 11, 2011, Tohoku earthquake was used to simulate this high-amplitude noise in the previous studies and serves as the best case of remote but not too far away sources of very high-amplitude signals at the IMS stations most important for the XSEL events – PETK, MKAR, KURK, and TXAR. A very important station PDYAR started in 2023 and did not participate in the Tohoku WCC processing.

For the current study, our principal objective is to estimate the capability of remote seismicity to affect the XSELs obtained for the days before and after the July 29, 2025, Kamchatka mega-earthquake. The set of 20 MEs was extended by 12 MEs covering the Tohoku aftershock zone, as presented in Figure 1. This allows for the investigation of the direct influence of the Tohoku aftershock sequence on the XSEL bulletins for the Kamchatka region and for the separate performance of the 20 Kamchatka MEs and 12 Tohoku MEs.

The first problem to address is how large the remote earthquakes' effect is on the Kamchatka XSELs generated with 20 MEs?  Furthermore, we must address whether we can suppress this effect using stochastic noise added to the actual waveforms? The period of the most intense aftershock activity between 07:00 and 13:00 UTC on March 11, 2011, was selected. The number of REB events in the aftershock zone of the Tohoku earthquake was a total of 232 with 173 of them having SEL3 seed events as the start points for the interactive review. There were no REB events within the studied Kamchatka region during these 6 hours of March 11, 2011.

The numbers of XSEL events in the twelve LA version-tolerance case configurations (indices 1 through 12) for the actual waveforms (*StN*=0.0) are shown in Table 4. These are the XSEL events found by the 20MEs within the Kamchatka region and *StN*=0.0 in the column "0.0" in Table 1. These are all new XSEL events as there were no REB events to match. The numbers decrease from 245 for index 1 to 2 for index 12. These are all ***false*** events wrongly created and located within the Kamchatka region from the WCC detections. The WCC detections were made by the 20 MEs and thus the WCC detection has relatively high side-sensitivity to the Kamchatka templates. For MKAR, KURK, and TXAR, this can be explained by the close station-event azimuths between the Kamchatka Peninsula and Japan; the southern border of the Kamchatka region is approximately 4° far away from the northern edge of the Tohoku zone as shown in Figure 1. Station PETK has arrivals from various azimuths from the 2025 Kamchatka aftershocks ranging from orthogonal to parallel to the direction of the Tohoku epicentre. Therefore, the side-sensitivity of the Kamchatka MEs to the Tohoku earthquakes is a natural result of their relative position with respect to the IMS stations.

The MEs near the southern edge along the 45° parallel should be the most sensitive to the creation of false events from valid detections. As an alternative, a sensitive ME can be deep and have short templates with lower specificity. Table 5 presents the numbers of XSEL events generated for two indices 1 and 7. These indices correspond to the weak and strict LA versions with a 5.0 s origin time tolerance which is the most prone to the creation of false events from valid detections. The most effective masters are indicated in bold. They are all close to the southern border of the studied region. The only exclusion is a deep event indicated in bold italic.

Table 4. Dependence of the number of XSEL events on *StN* for Tohoku earthquake found by the 20 MEs used for Kamchatka region.

| index | 0.00 | 0.005 | 0.01 | 0.05 | 0.10 | 0.25 | 0.50 | 1.00 | 2.50 | 5.00 |
|---|---|---|---|---|---|---|---|---|---|---|
| 1 | 235 | 238 | 264 | 312 | 298 | 243 | 161 | 90 | 27 | 0 |
| 2 | 234 | 220 | 246 | 314 | 287 | 228 | 150 | 81 | 23 | 0 |
| 3 | 197 | 197 | 224 | 300 | 263 | 205 | 141 | 71 | 21 | 0 |
| 4 | 137 | 140 | 174 | 265 | 235 | 166 | 107 | 47 | 13 | 0 |
| 5 | 102 | 90 | 121 | 187 | 179 | 124 | 82 | 28 | 7 | 0 |
| 6 | 63 | 54 | 68 | 132 | 118 | 76 | 58 | 21 | 5 | 0 |
| 7 | 43 | 50 | 77 | 148 | 149 | 103 | 68 | 31 | 5 | 0 |
| 8 | 33 | 35 | 59 | 129 | 113 | 84 | 55 | 25 | 5 | 0 |
| 9 | 24 | 24 | 40 | 105 | 104 | 67 | 44 | 18 | 4 | 0 |
| 10 | 9 | 15 | 29 | 77 | 72 | 39 | 23 | 13 | 2 | 0 |
| 11 | 4 | 5 | 12 | 38 | 37 | 22 | 12 | 4 | 1 | 0 |
| 12 | 2 | 3 | 5 | 18 | 17 | 10 | 4 | 1 | 0 | 0 |

Table 5. The number of XSEL events per one ME for index 1 and index 7 for the six-hour period on March 11, 2011.

| orid | 5050040 | 5376373 | 7220668 | **11231455** | ***12618922*** | **12688414** | **14686581** | 16007199 | 16313833 | **16438469** | 17328560 | 18425349 | **18517208** | 18683607 | 20280472 | **21227220** | 22605238 | 23316512 | 23329683 |
|---|---|---|---|---|---|---|---|---|---|---|---|---|---|---|---|---|---|---|---|
| index 1 | 1 | 5 | 1 | 8 | 21 | 45 | 20 | 3 | 9 | 34 | 6 | 4 | 30 | 4 | 12 | 7 | 1 | 12 | 5 |
| index 7 | 0 | 0 | 0 | 2 | 6 | 10 | 2 | 0 | 0 | 4 | 0 | 0 | 12 | 0 | 1 | 3 | 0 | 0 | 0 |

The other nine columns in Table 4 illustrate the change in the numbers of XSEL events for the twelve indices as a function of *StN*. There is a range of *StN* values where these numbers increase relative to the actual waveforms case in line with the findings reported in [Adushkin *et al.*, 2025; Kitov *et al.*, 2026; Kitov, 2026c]. The *StN* value maximizing the XSEL numbers for all indices likely resides between 0.05 and 0.1. This is an important observation for the WCC processing of the 2025 Kamchatka earthquake. The results presented in [Kitov, 2026d] can be improved by the addition of stochastic noise with scaled amplitude of approximately 0.075.

The largest *StN* value of 5.0 completely suppresses the generation of false events for all twelve indices. Unfortunately, this value also prohibits the creation of valid XSEL events in the WCC processing of the July 2025 data. The addition of stochastic noise is not helpful in the solution of the problem with the side-sensitivity of the WCC applied to IMS arrays. There is an indication of how to solve that problem, however, illustrated in Table 5. The events on the border of the studied Kamchatka region generate most of the false events. This effect does not significantly influence the estimates of the parameters important for the earthquake prediction as the MEs at the rim of the region should be excluded from the XSEL dataset related to these precursory variables.

Therefore, the use of the MEs within the Tohoku region would resolve the issue of false XSEL event hypotheses. These MEs should create valid XSEL events around their geographical positions from the valid detections at the same stations. Figure 1 shows the relative position of the twelve MEs, which cover the whole Tohoku region. This is a low number for the WCC processing to achieve high resolution for the Tohoku aftershocks, especially in the period of the most intense activity. These events are much more appropriate for the study of the Tohoku seismicity than any of the MEs used for the Kamchatka region.

The set of 32 MEs was used to process the same six-hour-long time interval with the region 35°N-65°N, 140°E-165°E, as shown in Figure 1. Table 6 presents the result of this processing. For *StN*=0.0, the total number of XSEL events for the set of 32 MEs decreases with index as for the other configurations in this study and as in [Kitov, 2025bd]. The number of matched REB events slightly grows from 113 for index 1 to 119 for index 3. The number of new XSEL events follows the decreasing trend of the total number of events. The partial result for the set of 20 Kamchatka MEs out of the full set of 32 MEs demonstrate non-zero values, i.e. the Kamchatka MEs found the Tohoku REB events. This possibility cannot be ruled out for the northernmost Tohoku aftershocks and high-quality Kamchatka MEs. For *StN*=0.05, which is the best value for XSEL event generation using the added stochastic noise in Table 4, the number of XSEL events is lower, but the number of matched REB events is larger than for *StN*=0.0. The number of false REB matches drops to the range between 0 and 2, and the number of new XSEL events falls within a range of 0 to 5. These new XSEL events were found in the high-amplitude Tohoku noise and are likely valid earthquakes within the Kamchatka region.

Table 6. Result of the WCC processing with 32 MEs of the 6h interval on March 11, 2011.

| | StN=0.0 | | | | | StN=0.05 | | | | |
|---|---|---|---|---|---|---|---|---|---|---|
| index | 32 MEs | | | 20 MEs | | 32 MEs | | | 20 MEs | |
| | Total | REB | New | REB | NEW | Total | REB | New | REB | NEW |
| 1 | 152 | 113 | 39 | 2 | 33 | 144 | 133 | 11 | 0 | 5 |
| 2 | 148 | 115 | 33 | 1 | 25 | 147 | 137 | 10 | 0 | 3 |
| 3 | 147 | 119 | 28 | 2 | 21 | 152 | 140 | 12 | 1 | 4 |
| 4 | 137 | 110 | 27 | 2 | 19 | 155 | 143 | 11 | 1 | 2 |
| 5 | 126 | 104 | 22 | 4 | 15 | 151 | 139 | 12 | 0 | 3 |
| 6 | 113 | 95 | 18 | 1 | 12 | 161 | 154 | 7 | 0 | 1 |
| 7 | 69 | 62 | 7 | 1 | 4 | 103 | 100 | 3 | 1 | 3 |
| 8 | 69 | 63 | 6 | 3 | 3 | 108 | 102 | 6 | 1 | 4 |
| 9 | 66 | 63 | 3 | 2 | 1 | 105 | 104 | 1 | 1 | 0 |
| 10 | 56 | 53 | 3 | 2 | 1 | 111 | 105 | 6 | 0 | 4 |
| 11 | 47 | 46 | 1 | 1 | 0 | 108 | 101 | 7 | 2 | 3 |
| 12 | 44 | 43 | 1 | 0 | 0 | 103 | 98 | 5 | 0 | 1 |

It is worth noting that the total number of XSEL events within the Tohoku region created by 12 MEs is lower than the total number of the false events created by the 20 MES. The CR process worked effectively to suppress the association of valid detections with false events. Additionally, station PETK, which is the most efficient for the Kamchatka region, has mediocre statistics for the Tohoku aftershocks. The station weights in the LA and CR process must be different for the 12 MEs. Nevertheless, even in a suboptimal setting of the WCC processing, the 12 MEs resolve the problem of IMS station side-sensitivity. In the prototype WCC pipeline, which worked several years in a test mode at the IDC, there were approximately 42,000 MEs optimally tuned to the localized station weights. There was no problem with the side-sensitivity as such with the dense covering of the whole zones of seismic activity. This problem was practically solved in [Kitov, 2026d] by the extension of the zone of the earthquake preparation to a broader region covering all seismic areas around the Kamchatka Peninsula.

## Discussion

The progress in seismological observations in the 20th century allowed scientists to obtain a large dataset of various seismic signals together with the estimates of their kinematic and dynamic parameters. The travel time curves were improved by the introduction of 3-D velocity models [Dziewonski, 1984; Tromp *et al*., 2005; Shapiro *et al*., 2010]. The amplitude-distance curves for primary and secondary seismic phases are converted into various magnitude scales [Gutenberg, 1945; Gutenberg and Richter, 1956; Veith and Clawson, 1972; Vaněk *et al*., 1982; Granville *et al*., 2005]. Attenuation coefficients and Q-factors are used to describe nonlinear mechanical processes in the Earth. Dispersion curves serve to understand the Earth’s fine elastic and inelastic properties. There is one common feature of all these parameters – extended statistics and well-defined uncertainty bounds. For example, the IDC does not use actual travel time residuals in the location algorithm [Coyne *et al*., 2012]. Instead, the theoretical travel time uncertainties are used, which are also dependent on the signal SNR value. These theoretical uncertainties are based on those in the travel time model ak135 [Kennett *et al*., 1995].

The WCC processing lacks such extensive datasets and accurate estimates of the principal parameters for the detected signals. In part, this is due to the extremely low SNR values, often below 1.0, of these WCC signals. The main reason, however, is the sporadic character of seismic studies based on the WCC processing. There is no global dataset to estimate the uncertainty of the travel time residuals and signal amplitudes to create a reference model allowing for the statistical assessments. One can make such estimates for the WCC signals that are also detected by standard methods and extrapolate the results to weaker signals. This is a promising start but there is no guarantee that the statistics for the weaker signals will follow the estimates for the visible seismic signals.

The value added by the addition of at least the same number of (low-magnitude) events to the global dataset as they contain now is worth the effort to reprocess the available raw waveform data. Small-scale experiments were conducted at the IDC with interactive review of the (automatic) XSEL bulletins in a few areas: China [Bobrov *et al*., 2014], Sumatra [Bobrov *et al*., 2016ab], Northern Atlantic [Bobrov *et al*., 2017]. From the XSEL seed events, IDC analysts added from 60% to 100% of newly created REB events. These are the events with visible signals and there were many more events not matching the analysts’ “visibility” experience. These exercises were important for the WCC results for the May 24, 2013, Sea of Okhotsk deep megaearthquake. There were no REB events detected prior to this event by standard methods, but hundreds of low-magnitude seismic events were detected by the WCC-based pipeline. The statistical significance of these XSEL events is supported by the results of the current study.

Comparison with the random noise waveforms is an attempt to estimate the reliability of the event hypotheses obtained in the WCC processing relative to the most basic case. This effort is important for the understanding of the WCC processing as such since it is a version of the matched filter method relying on stochastic noise conditions. For the purpose of the prediction of the May 24, 2013, Sea of Okhotsk and the July 29, 2025, Kamchatka earthquakes, which are based on the evolution of low-magnitude, and, thus, invisible to standard processing events, the estimates of statistical significance of the WCC-based hypotheses, are a mandatory step in the overall methodology. Any source of disturbance in the WCC processing has to be formally addressed and investigated in detail. The same applies to the side-sensitivity of the WCC at array stations. The Global Grid of master events [Kitov *et al*., 2016] practically dismisses this source of false events, but extended observations are needed to make this assumption measurable.

The operational constraints and stability metrics established in our numerical experiment provide the necessary methodological foundation for a targeted precursory wavefield analysis. It was demonstrated that the automated WCC pipeline is statistically protected against the random generation of XSEL event hypotheses under the optimized set of the origin-time tolerances. This allows the isolation of genuine tectonic activation patterns from baseline seismic field fluctuations related to a large number of seismic events with various magnitudes occurring at different distances from the studied area. The statistical significance of the XSEL events used to calculate the parameters for the precursory variables in [Kitov, 2026bd] is confirmed at a high level of confidence, as well as the statistical power of the XSEL events matching the REB during the periods of extremely high post-seismic activity.

The results for the first six hours of July 30 confirmed a saturation limit in the automated processing revealed in [Kitov, 2026d]. During this period of the highest seismic activity, the total number of XSEL events is almost independent of the origin time tolerance window. This behaviour is related to the rules governing the minimum allowable spacing between subsequent arrivals in the WCC detection process. When the aftershock activity is extremely intense, the smaller events are almost fully suppressed by the pipeline requirements. In this case, the total number of events remains constant between different tolerance cases for a given version.

The stochastic time series was generated by the *ran3* Fortran program. The random noise waveforms were added to the actual data after they were scaled to the maximum amplitude of the waveform in a given interval. The scaling factor *StN* was varied in a wide range from 0.001 to 100. For the largest *StN* value the random noise time series completely suppressed the actual waveform and the results of the WCC processing demonstrate the random generation of the XSEL event hypotheses. For the lower *StN* values, the efficiency of the WCC detector increased in line with the results of the previous studies [Adushkin *et al*., 2025; Kitov, 2026c; Kitov *et al*., 2026].

A grid search in the *StN* range from 0.0 to 1.0 allowed the estimation of the optimal value of *StN*=0.075 that maximizes the number of valid detections, XSEL events, and the match rate of the REB events. This is the closest point to the perfect matched filter conditions – the sought signal does not lose similarity with the template and the noise is as random as possible considering further degradation in the shape of the sought signal. This effect is similar to noise "whitening", which makes the matched-filter detector more sensitive to low-magnitude events. These events have an SNRcc deep below the routine IDC thresholds and are usually not visible to analysts. Therefore, the *StN*=0.075 configuration may help to improve the estimates of the precursory variables from the same actual data.

The determination of the optimal stochastic noise scaling factor—parametrically fixed at *StN*=0.075 — enables a high-resolution reprocessing of the continuous waveforms immediately preceding the megathrust failure. In the subsequent study, this calibrated *StN*=0.075 configuration will be directly deployed to investigate the evolution of low-magnitude seismicity prior to the July 29, 2025, Kamchatka megaearthquake. Specifically, we reevaluate the predictive power of the weak-to-strict XSEL event ratios as a linear function of magnitude scaling, establishing a robust empirical framework for intermediate-term earthquake forecasting within subduction zones.

The simulation with the 2011 Tohoku earthquake confirms that the side-sensitivity of the Kamchatka MEs is a natural result of their relative position to the IMS stations. High-amplitude regular phases from remote sources can travel along similar azimuths and generate valid WCC detections at stations PETK, MKAR, and KURK, which are then associated with false XSEL hypotheses. These false hypotheses are located within the Kamchatka region with a significant bias in hypocenter. The processing with the full set of 32 MEs demonstrates that the addition of 12 Tohoku MEs completely resolves this issue because the Tohoku MEs win the conflict resolution process. This confirms that the side-sensitivity problem can be solved by the extension of the master events list to a broader region covering adjacent seismic zones.

**References**

Adushkin, V. V., Kitov, I. O., and Sanina, I. A. (2025). Further development of the matched filter method for solving seismological problems. Doklady Earth Sciences, 523(1), 13–21. https://doi.org/10.1134/S1028334X25606182

Arrowsmith, S. J., and Eisner, L. (2006). A technique for identifying microseismic multiplets and application to the Valhall field, North Sea. Geophysics, 71(2), Q31–Q40. https://doi.org

Baisch, S., Ceranna, L., and Harjes, H.-P. (2008). Earthquake clusters: What can we learn from waveform similarity? Bulletin of the Seismological Society of America, 98(6), 2806–2814. https://doi.org

Bobrov, D., Kitov, I., and Zerbo, L. (2014). Perspectives of cross-correlation in seismic monitoring at the International Data Centre. Pure and Applied Geophysics, 171(3–5), 439–468. https://doi.org

Bobrov, D. I., Kitov, I. O., Rozhkov, M. V., and Friberg, P. (2016a). Towards global seismic monitoring of underground nuclear explosions using waveform cross-correlation. Part I: Grand master events. Seismic Instruments, 52(1), 43–59. https://doi.org

Bobrov, D. I., Kitov, I. O., Rozhkov, M. V., and Friberg, P. (2016b). Towards global seismic monitoring of underground nuclear explosions using waveform cross-correlation. Part II: Synthetic master events. Seismic Instruments, 52(3), 207–223. https://doi.org

Bobrov, D., Kitov, I., and Rozhkov, M. (2017). Studying seismicity of the Atlantic Ocean using waveform cross-correlation. NNC RK Bulletin, 2(70), 5–19.

Comprehensive Nuclear-Test-Ban Treaty (1996). Protocol to the Comprehensive Nuclear-Test-Ban Treaty. https://ctbto.org

Coyne, J., Bobrov, D., Bormann, P., Duran, E., Grenard, P., Haralabus, G., Kitov, I., and Starovoit, Yu. (2012). Chapter 15: CTBTO: Goals, networks, data analysis and data availability. In: P. Bormann (Ed.), New Manual of Seismological Practice (NMSOP-2), GFZ Potsdam, 1–41. https://doi.org

Dziewonski, A. M. (1984). Mapping the lower mantle: Determination of lateral heterogeneity in P velocity up to degree and order 6. Journal of Geophysical Research, 89(B7), 5929–5952. https://doi.org/10.1029/JB089iB07p05929

Granville, J. P., Richards, P. G., Kim, W.-Y., and Sykes, L. R. (2005). Understanding the differences between three teleseismic mb scales. Bulletin of the Seismological Society of America, 95(5), 1809–1824. https://doi.org

Gutenberg, B. (1945). Amplitudes of P, PP, and S and magnitude of shallow earthquakes. Bulletin of the Seismological Society of America, 35(2), 57–69. https://doi.org

Gutenberg, B., and Richter, C. F. (1956). Magnitude and energy of earthquakes. Annali di Geofisica, 9(1), 1–15. https://doi.org

Israelsson, H. (1990). Correlation of waveforms from closely spaced regional events. Bulletin of the Seismological Society of America, 80(6), 2177–2193.

Joswig, M. (1990). Pattern recognition for earthquake detection. Bulletin of the Seismological Society of America, 80(1), 170–186.

Kennett, B. L. N., Engdahl, E. R., and Buland, R. (1995). Constraints on seismic velocities in the Earth from travel times. Geophysical Journal International, 122(1), 108–124. https://doi.org

Kitov, I. O. (2026a). Low-magnitude seismic activity between the Kamchatka July 20 and July 29, 2025, earthquakes. Spatio-temporal evolution recovered using waveform cross-correlation. arXiv:2601.15302. https://doi.org/10.48550/arXiv.2601.15302

Kitov, I. O. (2026b). Spatio-temporal evolution of low-magnitude seismicity before the May 24, 2013, Sea of Okhotsk earthquake recovered by waveform cross-correlation. Is it an earthquake prediction case? arXiv:2603.26717. https://doi.org/10.48550/arXiv.2603.26717

Kitov, I. O. (2026c). Seismic noise suppression: array stations, waveform cross-correlation, and noise stochastization. arXiv:2604.21939. https://doi.org/10.48550/arXiv.2604.21939

Kitov, I., Bobrov, D., and Rozhkov, M. (2016). Automatic Event Bulletin Built by Waveform Cross-Correlation using the Global Grid of Master Events with Adjustable Templates. Geophysical Research Abstracts, Vol. 18, EGU2016-6478, EGU General Assembly 2016.

Kitov, I. O. (2026d). Prediction of the Kamchatka July 29, 2025, earthquake by the evolution of low-magnitude seismicity recovered using waveform cross-correlation at IMS seismic arrays. arXiv:2606.17060. https://doi.org/10.48550/arXiv.2606.17060

Kitov, I. O., Sanina, I. A., Sokolova, I. N., and Vinogradov, Yu. A. (2026). Study of Seismic Activity between the Mainshock and the First Aftershock of the Kamchatka Earthquake on July 29, 2025. Russian Journal of Earth Sciences (in press).

Press, W., Teukolsky, S., Vetterling, W., and Flannery, D. (1986). Numerical Recipes in Fortran 77: The Art of Scientific Computing, Volume 1. Cambridge University Press.

Saragiotis, C., and Kitov, I. (2020). Tuning IMS station processing parameters and detection thresholds to increase detection precision and decrease detection miss rate. EGU General Assembly 2020, Online, 4–8 May 2020, EGU2020-8949. https://doi.org

Schaff, D. P., and Richards, P. G. (2004). Repeating seismic events in China. Science, 303(5661), 1176–1178. https://doi.org

Schaff, D. P., Kim, W.-Y., and Richards, P. G. (2025). Background Seismicity for parts of the northern Korean peninsula. CTBT Science & Technology Conference, P1.2-714, Vienna, Austria.

Schweitzer, J., Fyen, J., Mykkeltveit, S., Gibbons, S. J., Pirli, M., Kühn, D., and Kværna, T. (2012). Seismic arrays. In: P. Bormann (Ed.), New Manual of Seismological Practice (NMSOP-2), GFZ Potsdam, Ch. 9, 1–45. https://doi.org

Selby, N. (2010). Relative location of the October 2006 and May 2009 DPRK announced nuclear tests using International Monitoring System Seismometer arrays. Bulletin of the Seismological Society of America, 100(4), 1779–1784. https://doi.org

Shapiro, N. M., Campillo, M., Stehly, L., and Ritzwoller, M. H. (2005). High-resolution surface-wave tomography from ambient seismic noise. Science, 307(5715), 1615–1618. https://doi.org

Tromp, J., Tape, C., and Liu, Q. (2005). Seismic tomography, adjoint methods, time reversal and banana-doughnut kernels. Geophysical Journal International, 160(1), 195–216. https://doi.org/10.1111/j.1365-246X.2004.02453.x

Turin, G. L. (1960). An introduction to matched filters. IRE Transactions on Information Theory, 6(3), 311–329. https://doi.org

Vaněk, J., Kondorskaya, N. V., Fedorova, I. V., and Christoskov, L. (1982). Optimization of amplitude curves of seismic P-, S- and L-waves in the homogeneous magnitude system of the Eurasian continent. Tectonophysics, 84(1), 41–45. https://doi.org

Veith, K. F., and Clawson, G. E. (1972). Magnitude from short-period P-wave data. Bulletin of the Seismological Society of America, 62(2), 435–452. https://doi.org

Waldhauser, F., and Schaff, D. (2008). Large-scale cross-correlation-based relocation of two decades of northern California seismicity. Journal of Geophysical Research, 113(B8), B08311. https://doi.org

Appendix 1. IDC parameters for 32 MEs.

| IDC orid | origin time, s | Lat, deg | Lon, deg | Nass* | $m_b$ | depth, km |
|---|---|---|---|---|---|---|
| 23689063 | 1678291851.10 | 51.621 | 159.579 | 54 | 4.9 | 12 |
| 23329683 | 1671934664.20 | 56.12 | 160.994 | 90 | 4.5 | 115.7 |
| 23316512 | 1671816197.60 | 54.894 | 159.942 | 83 | 4.1 | 116.1 |
| 22605238 | 1659376874.20 | 50.535 | 150.466 | 65 | 3.8 | 478.2 |
| 21227220 | 1633993802.20 | 48.281 | 153.773 | 90 | 5.1 | 87.3 |
| 20280472 | 1615919903.20 | 54.786 | 162.806 | 91 | 5.9 | 21.8 |
| 18683607 | 1585166167.90 | 55.894 | 160.257 | 109 | 4.7 | 190.4 |
| 18517208 | 1581590025.80 | 45.758 | 148.842 | 96 | 6 | 155.6 |
| 18425349 | 1579691054.10 | 54.918 | 161.407 | 119 | 5 | 67.2 |
| 17328560 | 1557210493.60 | 49.518 | 155.738 | 78 | 4.7 | 55 |
| 16438469 | 1539380306.30 | 47.197 | 146.587 | 108 | 4.4 | 373.7 |
| 16313833 | 1536524928.20 | 52.296 | 157.106 | 77 | 4 | 146.8 |
| 16007199 | 1530841206.30 | 51.57 | 157.753 | 75 | 5.4 | 64.4 |
| 14686581 | 1501448448.90 | 46.179 | 150.959 | 101 | 5.5 | 89.4 |
| 12688414 | 1450387899.40 | 47.935 | 146.905 | 143 | 4.7 | 449.5 |
| 12618922 | 1448704593.10 | 52.677 | 152.704 | 95 | 4.2 | 527.2 |
| 11231455 | 1410892779.90 | 45.125 | 147.038 | 51 | 4.7 | 0 |
| 7220668 | 1299285065.70 | 52.798 | 160.797 | 63 | 4.7 | 0 |
| 5376373 | 1241817750.30 | 58.084 | 164.32 | 36 | 5 | 12.1 |
| 5050040 | 1227552258.00 | 53.029 | 159.581 | 64 | 4.4 | 43.8 |
| 21963352 | 1647441392.94 | 37.71 | 141.55 | 103 | 6 | 59.4 |
| 17806771 | 1567036001.90 | 40.99 | 143.00 | 94 | 5.4 | 42.1 |
| 17236058 | 1554970703.24 | 40.36 | 143.35 | 93 | 5.3 | 35.8 |
| 17078267 | 1551900361.07 | 38.7 | 141.60 | 74 | 4.5 | 67.9 |
| 13552398 | 1473421997.68 | 36.36 | 140.98 | 71 | 4.7 | 51.9 |
| 10997485 | 1404513724.83 | 39.67 | 142.00 | 69 | 5.4 | 47.5 |
| 9669379 | 1365192052.13 | 36.72 | 141.34 | 118 | 4.8 | 46.2 |
| 9001616 | 1346267113.19 | 38.39 | 141.84 | 116 | 5.2 | 68.1 |
| 7363085 | 1300221003.01 | 35.21 | 141.06 | 56 | 5.1 | 42.3 |
| 7360668 | 1300182590.91 | 37.32 | 142.42 | 80 | 5.5 | 0 |
| 7336822 | 1299838230.51 | 39.24 | 142.77 | 88 | 5.6 | 0 |
| 7297161 | 1299899521.09 | 35.94 | 141.44 | 77 | 5.2 | 41.3 |
| | | | | | | |

*Nass – number of associated phases

Appendix 2. Principal Event Definition Criteria. Cited [Kitov, 2026d].

"The approach was simplified by the introduction of two LA versions, each with different sets of defining parameters and six cases per version allowing for the fine-tuning of statistical significance. These two versions needed to be sufficiently distinct in a statistical sense to avoid overlap between the cases. The strict version of the LA parameters has to be close to the corner magnitude of the recurrence curve for the region in order to provide a smooth transition from the XSEL to the REB. This version should not generate any XSEL events during quiet seismic periods, similar to the REB. The following LA parameters are used for this task: 1) the number of associated stations is 3. 2) The origin-time tolerance is a case dependent parameter. 3) The minimum total event weight is 2.5. 4) The weight of one of the best stations to be associated with any 3- to 5-station events is 0.855. 5) The minimum sums of SNRcc values for 3-, 4-, and 5-station events are 15.0, 18.5, and 22.0, respectively. 6) The lowest possible value for the minimum SNRcc at one of the top stations from 3) is 5.0. 7) The grid radius is 48 km, defined by 12 steps of 4 km. The event hypotheses beyond a radius of 43.2 km are rejected. This radius can be less than half the distance between neighboring MEs, but it is important to increase statistical significance by reducing the flexibility in location.

On the opposite side of the XSEL sensitivity is the weak LA version. The defining parameters were guesstimated using the experience with LA processing in a number of previous studies and from the SNRcc curves in Figures 6 through 8, as well as similar curves for other involved stations. The parameters for the weak LA version are as follows: 1) the number of associated stations is 3. 2) The origin-time tolerance is a case dependent parameter. 3) The minimum total event weight is set at 1.8. 4) The weight of one of the top stations to be associated with any 3- to 5-station events is 0.80. 5) The minimum sums of SNRcc values for 3-, 4-, and 5-station events are 14.0, 17.5, and 21.0, respectively. 6) The lowest possible value for the minimum SNRcc at one of the top stations from 3) is 4.5. 7) The grid radius is 90 km, defined by 15 steps of 6 km. The event hypotheses beyond a radius of 81.0 km are rejected.

After a brief test, six different origin-time tolerances, Δt, were introduced to determine their impact on the XSELs from the same detection lists: 5.0, 3.0, 2.0, 1.0, 0.5, and 0.25 seconds. These six tolerances create 12 version/case pairs. They can be aligned in a formal order: $v_1c_1$, ..., $v_1c_6$, $v_2c_1$,...,$v_2c_6$, where $v_1c_1$ corresponds to the weak LA version with $\Delta t$ =5.0 s. The shortest $\Delta t$ allows for practically only actual event hypotheses to be created. For detections randomly distributed in time and in SNRcc value, the probability of 3 or more of them at the best stations having origin times in a 0.5 s time window is extremely low, considering the observed detection rates of 30 to 40 per hour. The downside of having a shorter $\Delta t$ is the higher likelihood of missing many weaker but valid XSEL event hypotheses, associating weaker signals with poor arrival-time estimates and larger travel-time residuals. The $v_2c_6$ pair should not be able to detect too many new XSEL events in addition to the REB. Overall, the XSEL events that match the REB events have to be of a higher quality related to the SNRcc values of the associated signals.

The weak LA version with the narrowest origin-time window, $v_1c_6$, has to be focused on the new XSEL events with the best quality below the corner magnitude of the recurrence curve. These events are highly statistically significant, but can also include weaker WCC arrivals. The $v_1c_1$ pair has the highest resolution. It may contain many valid events and likely some false events, with both types formally matching the EDC. For an XSEL obtained with a given version and case pair, a formal statistical threshold can be defined to distinguishing between valid and false events as determined by the LA algorithm. Then, the strict and weak versions for the same case, $\Delta t$, define a specific range of event quality between their corresponding thresholds. The assumption behind introducing two versions and six cases was that the corner magnitude of the recurrence curve of the detected XSEL events depends on these thresholds of statistical significance. The potential influence of source mechanisms, as well as of the noise level at stations around the arrival times of all associated phases, is averaged as for the long-term REB recurrence curve."